\documentclass[aip,twocolumn,reprint,superscriptaddress]{revtex4-1}
\usepackage{graphicx}
\usepackage{amsmath,amssymb,mathtools,bbold} 
\usepackage{color}
\usepackage[colorlinks=true,linkcolor=blue,citecolor=blue]{hyperref}
\usepackage{stix}

\newcommand{\revone}[1]{\textcolor{black}{#1}}

\begin{document}

\title{Analysis of the singular band structure occurring in one-dimensional topological normal and superfluid fermionic systems: A  pedagogical description}

\author{Marcello Calvanese Strinati}
\affiliation{Centro Ricerche Enrico Fermi (CREF), Via Panisperna 89a, 00184 Rome, Italy}
\author{Giancarlo Calvanese Strinati}
\affiliation{School of Science and Technology, Physics Division, Universit\`{a} di Camerino, 62032 Camerino, Italy}

\noindent
\date{\today}

\begin{abstract}
Topological properties of solid-state materials arise when crossings occur in their band-structure eigenvalues, which give rise to discontinuities in the associated Bloch-function eigenvectors once these are mapped over the whole Brillouin zone.
\revone{These non-analytic properties have direct consequences on the spatial decay of the corresponding Wannier functions, leading to what is nowadays referred to as the ``obstruction to finding symmetric Wannier functions'' for a given set of bands, as well on the need for shifting the Wannier functions to interstitial positions, related to what is nowadays known as the ``bulk-boundary correspondence''. The importance of non-analytic points of Bloch eigenfunctions and their consequences  for the spatial decay of Wannier functions were historicallly anticipated back in 1978 [G. Strinati, Phys. Rev. B \textbf{18}, 4104-4119 (1978)], somewhat before the work of Berry on what came to be referred to as the ``Berry phase'' [M. V. Berry, Proc. R. Soc. London A \textbf{392}, 45 (1984)]. In particular, the former paper identified key precursors and physical insights that are now understood, in hindsight, to be closely related to the later developments mentioned above.}
Here, we recap the essential features of these key issues in a rather pedagogical way, by considering in full details two instructing examples for which the origin of the discontinuities in the eigenvectors can be readily traced and mapped out, and the rate of the spatial falloff of the associated Wannier functions can be fully determined. For this analysis to be as complete as possible, two cases, one for non-interacting and one for interacting fermions, are considered on equal footing.
\end{abstract}

\maketitle

\section{Introduction} 
\label{sec:introduction}
The geometric properties of Bloch states in the Brillouin zone are manifestation for condensed-matter systems of the broad concept of Quantum Geometry \cite{Torma-2023,Bernevig-Torma-2025}. 
Although the associated topology can be formally encoded in general concepts like the Berry curvature, the quantum geometric tensor, the Chern number, and so on \cite{Bernevig-2013,Shen-2017,Vanderbilt-2018},
in the ultimate analysis it all boils down to the occurrence of discontinuities in the Bloch-function eigenvectors as mentioned above.
It thus appears worthwhile to provide an expanded and pedagogical description of these discontinuities in simple cases, which to a large extent can be described analytically and require minimal numerical work.

In Ref.~\cite{Strinati-1978} the discontinuities of the Bloch-function eigenvectors were \revone{brought to light} while following with continuity the evolution of the Bloch functions over the constant energy surfaces embedded in the Brillouin Zone (BZ) 
of a three-dimensional wave-vector space. 
In the present article, we analyse a simpler situation and follow the evolution of the Bloch functions in a one-dimensional wave-vector space, by considering two model systems for spin-polarized and spin-less fermions which share a similar behavior, namely:
(i) a p-wave fermionic superfluid where the effects of the inter-particle interaction are treated at the mean-field level, and 
(ii) a non-interacting (normal) two-leg ladder model with orbitals of different symmetry which exemplifies a topological insulator.
In both cases, the von Neumann-Wigner theorem \cite{vN_W-1929}, which forbids band crossing in strictly one spatial dimension, is circumvented either by the presence of the additional particle-hole symmetry in the superfluid case (i)
or by moving in ($1 \! + \! \epsilon$)-dimensional space in the normal case (ii).
In addition, in both cases the system parameters (specifically, the chemical potential) can be varied with continuity in such a way to pass from a fully topological to a topological trivial phase (and vice-versa) by crossing a quantum critical point (QCP) \cite{Hertz-1976}.

In Ref.~\cite{Strinati-1978} it was further anticipated that the discontinuities of the Bloch-function eigenvectors play a key role in determining the localization properties of the associated Wannier functions as well as their center of symmetry.
In the more modern language of topological insulators \cite{Calandra-2007,Vishwanath-2018}, one speaks \revone{accordingly} of an ``obstruction to finding symmetric Wannier functions”, meaning by that an obstruction occurs for finding a gauge where the Wannier functions 
(obtained by the Fourier transform of the Bloch wave functions) are exponentially localized in space.
In addition, the connection between Wannier centers and Berry phases has been found to be of fundamental importance \cite{Vanderbilt-2018}, being also related to the so-called ``bulk-boundary correspondence'' 
which is regarded to be the foundational principle of topological physics \cite{Vanderbilt-2018,Torma-2023}.
In the present article, a detailed analysis of this kind of obstruction will be provided for the Wannier functions associated with the bands of the topological insulator of case (ii) above, where the spatial falloff of the Wannier functions
will be seen to pass from an exponential behavior in the trivial topological phase to a power-law decay in the non-trivial topological phase upon crossing the QCP.
On the other hand, for the topological fermionic superfluid of case (i) above, a similar obstruction is again revealed in the relevant Wannier functions, while the Cooper-pair wave function, when referred to its value at large inter-particle separation, shows a behavior similar to that of 
correlation functions across the QCP.

In this way, we expect the present work to be of some help to non-specialist readers for getting a simple introduction to the topology of electronic band structure, while getting straight to the core of the subject.

\section{One-dimensional p-wave fermionic superfluid}
\label{sec:fermionic-superfluid}
We open by considering the topological properties of a p-wave fermionic superfluid in one spatial dimension.
Although this topic was already considered in some detail in Ref.~\cite{Neupert-2020} (where special emphasis was given to the Wannier obstruction in real space), for a fuller understanding of the topic we regard it important 
to expand further on that analysis, by providing also a detailed description of the change in the character of the band eigenvectors when crossing the QCP that separates the non-trivial from the trivial topological phases,
and giving at the same time a more detailed analysis of the associated Wannier obstruction.
In addition, differences will result between the present analysis and that of Ref.~\cite{Neupert-2020}, regarding the spatial behavior of the Cooper-pair wave function across the QCP.

The system we consider corresponds to a set of spin-polarized fermions embedded in a one-dimensional lattice with $N$ cells (with the lattice constant equal to one for convenience), described by the Hamiltonian $H = H_{1} + H_{2}$ with
\begin{eqnarray}
H_{1} & = & \int \! dx \, \Psi^{\dagger}(x) \, h(x) \, \Psi(x)
\label{system-Hamiltonian-noninteracting-1} \\
H_{2} & = & \int \! dx \, dx' \, \Psi^{\dagger}(x) \, \Psi^{\dagger}(x') \, V(x-x') \, \Psi(x') \, \Psi(x) \,\, ,
\label{system-Hamiltonian-interacting-1}
\end{eqnarray}
where $\Psi(x)$ is a field operator (for the given spin) and the single-particle Hamiltonian $h(x)$ contains the lattice potential.
On physical grounds, we also assume the presence of a ``ghost'' band at higher energy which can be partly occupied by opposite spin-polarized fermions, in such a way that the chemical potential can be varied sufficiently at will 
by the action of an external magnetic field (not explicitly considered).
\revone{This consideration is in analogy with what occurs for an intrinsic semiconductor, where at low temperature the chemical potential lies approximately halfway between the (occupied) valence band and the (empty) valence band~\cite{Ashcroftmermin1976}.}

\subsection{Diagonalization of the one-body Hamiltonian}
\label{sec:diagonalization}
We restrict to one-band approximation, by representing
\begin{equation}
\Psi(x) = \sum_{n} \phi(x-n) \, c_{n} 
\label{one-band-approximation}
\end{equation}
where $\phi(x-n) $ is a single-particle (real) orbital centered at site $n$ and $c_{n}$ is the corresponding annihilation operator. 
With the definition of the (positive definite) \emph{hopping integral\/}
\begin{equation}
t = - \int \! dx \, \phi(x-(n+1)) \, h \, \phi(x-n)
\label{hopping-integral}
\end{equation}
for nearest neighbors only, the one-body part $H_{1}$ of the Hamiltonian takes the form
\begin{equation}
H_{1} = - \, t \, \sum_{n} \left( c^{\dagger}_{n+1} c_{n} + c^{\dagger}_{n-1} c_{n} \right) \,\, .
\label{system-Hamiltonian-noninteracting-2}
\end{equation}
This Hamiltonian is diagonalized in reciprocal space, by introducing the lattice Fourier transform
\begin{equation}
c_{n} = \frac{1}{\sqrt{N}} \, \sum_{k}^{BZ} e^{i n k} c_{k} 
\label{lattice-Fourier-transform}
\end{equation}
where the wave vector $k \in (-\pi,+\pi)$ is restricted to the BZ, such that
\begin{equation}
\frac{1}{N} \, \sum_{k}^{BZ} e^{i n (k -  k')} = \delta_{k,k'} \, .
\label{lattice-delta-function}
\end{equation}
Equation~\eqref{system-Hamiltonian-noninteracting-2} then becomes
\begin{equation}
H_{1} = - 2 \, t \, \sum_{k}^{BZ} \, \cos(k) \, c^{\dagger}_{k} \, c_{k} = \sum_{k}^{BZ} \, \lambda(k) \, c^{\dagger}_{k} \, c_{k}
\label{system-Hamiltonian-noninteracting-3}
\end{equation}
where $\lambda(k) = - 2 \, t \, \cos(k)$ is the single-particle dispersion relation.

For later purposes, it is convenient to measure the single-particle energy levels from the chemical potential $\mu$, by defining
\begin{equation}
\xi(k) = \lambda(k) - \mu=- 2 \, t \, \cos(k)-\mu \,\, ,
\end{equation}
and write for the non-interacting part of the grand-canonical Hamiltonian
\begin{eqnarray}
& & \hspace{-0.5cm} H_{1} - \mu \, N = \sum_{k}^{BZ} \xi(k) \, c^{\dagger}_{k} \, c_{k} \nonumber\\
&=&  \frac{1}{2}\sum_{k}^{BZ} \left[\xi(k) \, c^{\dagger}_{k} \, c_{k}+\xi(-k) \, c^{\dagger}_{-k} \, c_{-k}\right]
\nonumber \\
& = & \frac{1}{2} \sum_{k}^{BZ} \left(\begin{array}{cc} \hspace{-0.1cm} c^{\dagger}_{k} & c_{-k} \hspace{-0.1cm} \end{array}\right) \!
\left( \hspace{-0.1cm} 
\begin{array}{cc}
\xi(k)     &         0  \\
0          & - \xi(-k)
\end{array}
\hspace{-0.1cm} \right) \! 
\left( \hspace{-0.1cm} 
\begin{array}{c}
c_{k}   \\
c^{\dagger}_{-k}     
\end{array}
\hspace{-0.1cm} \right) 
+  \frac{1}{2} \sum_{k}^{BZ} \xi(-k) \,\, ,\nonumber\\
\label{system-Hamiltonian-noninteracting-4}
\end{eqnarray}
where $N = \sum_{k}^{BZ} c^{\dagger}_{k} \, c_{k}$, $\xi(-k) = \xi(k)$ , and the anti-commutation relation $c^{\dagger}_{-k}c_{-k}=1-c_{-k}c^{\dagger}_{-k}$ has been used. 
Note that the chemical potential contains the effect of the on-site term $\int \! dx \, \phi(x) \, h \, \phi(x)$, which was purposely not included in the Hamiltonian~\eqref{system-Hamiltonian-noninteracting-2}.

\subsection{Mean-field approximation of the two-body Hamiltonian}
\label{sec:mean-field-Hamiltonian}
Next, we approximate the interacting part $H_{2}$ of the Hamiltonian by a \emph{mean-field\/} ($\mathrm{mf}$) \emph{pairing decoupling\/}. 
Using the representation (\ref{one-band-approximation}) of the field operator and considering nearest-neighbor interaction only, we obtain
\begin{equation}
H_{2}^{\mathrm{mf}} = \sum_{n} \left( \Delta \, c^{\dagger}_{n+1} c^{\dagger}_{n} + \Delta^{*} \, c_{n} c_{n+1} \right)
\label{system-Hamiltonian-interacting-2}
\end{equation}
where
\begin{equation}
\Delta = v_{0} \, \langle c_{n} c_{n+1} \rangle
\label{local-gap-definition}
\end{equation}
(taken real without loss of generality) is independent of $n$ due to translational invariance, with the \emph{negative\/} constant
\begin{equation}
v_{0} = 2 \int \! dx \, dx' \, |\phi(x-n)|^{2} \, V(x-x') \, |\phi(x'-(n-1))|^{2} \,\, .
\label{v_0-definition}
\end{equation}
Here and in the following,  $\left\langle \cdots \right\rangle$ stands for the (self-consistent) average over the ground state of the mean-field grand-canonical Hamiltonian (see below), 
since in the present article we consider the zero-temperature limit only. 
Note in addition that the quantity $\langle c_{n} c_{n+1} \rangle$ in Eq.~(\ref{local-gap-definition}) corresponds to a $p$-wave superconducting behavior (while the $s$-wave contribution $\langle c_{n} c_{n} \rangle$ vanishes for aligned spins). 
With the lattice Fourier transform (\ref{lattice-Fourier-transform}), the mean-field two-body Hamiltonian (\ref{system-Hamiltonian-interacting-2}) becomes
\begin{eqnarray}
H_{2}^{\mathrm{mf}} & = & \frac{1}{2} \sum_{k}^{BZ} \left( \Delta(k)\,c^{\dagger}_{k} c^{\dagger}_{-k} + \Delta^{*}(k)\,c_{-k} c_{k} \right)
\nonumber \\
& = & \frac{1}{2} \sum_{k}^{BZ} \left(\begin{array}{cc}c^{\dagger}_{k}&c_{-k}\end{array}\right) \!
\left( \hspace{-0.1cm} 
\begin{array}{cc}
0                    &    \Delta(k)   \\
\Delta^{*}(k)   &          0
\end{array}
\hspace{-0.1cm} \right) \! 
\left( \hspace{-0.1cm} 
\begin{array}{c}
c_{k}   \\
c^{\dagger}_{-k}     
\end{array}
\hspace{-0.1cm} \right) \,\, ,
\label{system-Hamiltonian-interacting-3}
\end{eqnarray}
where
\begin{equation}
\Delta(k) = - 2 \, i \, \Delta \, \sin (k) 
\label{k_dependent-gap-definition}
\end{equation}
with $\Delta$ defined in Eq.~(\ref{local-gap-definition}).
Taking again advantage of translational invariance and using the lattice Fourier transform \eqref{lattice-Fourier-transform}, the quantity $\Delta$ of Eq.~(\ref{local-gap-definition}) acquires the convenient form:
\begin{equation}
\Delta = \frac{v_{0}}{N} \, \sum_{n} \, \langle c_{n} c_{n+1} \rangle  = - i \, \frac{v_{0}}{N} \, \sum_{k}^{BZ} \, \sin (k) \, \langle c_{k} c_{-k} \rangle \, .
\label{local-gap-manipulated-1}
\end{equation}

\subsection{Diagonalization of the mean-field Hamiltonian}
\label{sec:mean-field-grand-canonical-Hamiltonian}	
\vspace{0.2cm}
By combining the expressions~\eqref{system-Hamiltonian-noninteracting-4} and~\eqref{system-Hamiltonian-interacting-3}, the complete mean-field grand-canonical Hamiltonian reads:
\begin{eqnarray}
H^{\mathrm{mf}} - \mu N 
& = & \frac{1}{2} \sum_{k}^{BZ} \left(\begin{array}{cc} \hspace{-0.1cm} c^{\dagger}_{k} & c_{-k} \hspace{-0.1cm} \end{array}\right) \!
\left( \hspace{-0.1cm} 
\begin{array}{cc}
\xi(k)               &     \Delta(k)  \\
\Delta^{*}(k)    &      - \xi(k)
\end{array}
\hspace{-0.1cm} \right) \! 
\left( \hspace{-0.1cm} 
\begin{array}{c}
c_{k}   \\
c^{\dagger}_{-k}     
\end{array}
\hspace{-0.1cm} \right) 
\nonumber \\
 & + & \frac{1}{2} \sum_{k}^{BZ} \xi(k) \, .
\label{mean-field-total-system-Hamiltonian-1} 
\end{eqnarray}
In this expression, diagonalization of  the matrix
\begin{equation}
M(k) = 
\left( \hspace{-0.1cm} 
\begin{array}{cc}
\xi(k)               &     \Delta(k)  \\
\Delta^{*}(k)    &       - \xi(k)
\end{array}
\hspace{-0.1cm} \right) \
\label{M-matrix}
\end{equation}
yields
\begin{equation}
\left( \hspace{-0.1cm} 
\begin{array}{cc}
\xi(k)               &     \Delta(k)  \\
\Delta^{*}(k)    &     - \xi(k)
\end{array}
\hspace{-0.1cm} \right) \! 
\left( \hspace{-0.1cm} 
\begin{array}{c}
a_{k}   \\
b_{k}     
\end{array}
\hspace{-0.1cm} \right)
= E(k) \! 
\left( \hspace{-0.1cm} 
\begin{array}{c}
a_{k}   \\
b_{k}     
\end{array}
\hspace{-0.1cm} \right) 
\label{diagonalization-M-matrix}
\end{equation}
where $E(k) = \pm \, \epsilon(k) $ with
\begin{equation}
\epsilon(k) = \sqrt{ \xi^2(k) + |\Delta(k)|^{2}} \,\, ,
\label{epsilon}
\end{equation}
and with the normalization condition
\begin{equation}
{|a_{k}|}^{2} + {|b_{k}|}^{2} = 1
\label{normalization-condition}
\end{equation}
for any $k$ in the BZ. 
In the two ($\pm$) cases, we obtain:

\vspace{0.1cm}
\noindent
(i)  + sign:  Upper band, with energy $E(k) = + \, \epsilon(k) $ and eigenvector components
\begin{subequations}
\begin{align}
a_{k}^{(+)} &= \sqrt{\frac{1}{2} \, \left( 1 + \frac{\xi(k)}{\epsilon(k)} \right)} \label{a-plus-sign} \\
b_{k}^{(+)} & = \frac{\Delta^*(k)}{|\Delta(k)|} \, \sqrt{\frac{1}{2} \, \left( 1 - \frac{\xi(k)}{\epsilon(k)} \right)} \label{b-plus-sign} \,\, ;
\end{align}
\label{eq:bandplusminusvector01}
\end{subequations}

\vspace{0.1cm}
\noindent
(ii)  - sign:  Lower band, with energy $E(k) = - \, \epsilon(k) $ and eigenvector components
\begin{subequations}
\begin{align}
a_{k}^{(-)} &= \sqrt{\frac{1}{2} \, \left( 1 - \frac{\xi(k)}{\epsilon(k)} \right)} \label{a-minus-sign} \\
b_{k}^{(-)} &= - \, \frac{\Delta^*(k)}{|\Delta(k)|} \, \sqrt{\frac{1}{2} \, \left( 1 + \frac{\xi(k)}{\epsilon(k)} \right)} \label{b-minus-sign} \,\, .
\end{align}
\label{eq:bandplusminusvector02}
\end{subequations}

\noindent
In the expressions (\ref{eq:bandplusminusvector01}) and (\ref{eq:bandplusminusvector02}), $\Delta^*(k)/|\Delta(k)| = i \, {\rm sign} (k) $ for real $\Delta$. 
We can further group together the results~\eqref{eq:bandplusminusvector01} and~\eqref{eq:bandplusminusvector02} by introducing the notation
\begin{subequations}
\begin{align}
u(k) &= \sqrt{\frac{1}{2} \, \left( 1 + \frac{\xi(k)}{\epsilon(k)} \right)}
\label{u-notation} \\
v(k) &= i \, {\rm sign} (k) \, \sqrt{\frac{1}{2} \, \left( 1 - \frac{\xi(k)}{\epsilon(k)} \right)} \,\, ,
\label{v-notation}
\end{align}
\label{eq:uvnotation01}
\end{subequations}
such that Eqs.~\eqref{eq:bandplusminusvector01} and~\eqref{eq:bandplusminusvector02} can be cast in the compact form
\begin{subequations}
\begin{align}
a_{k}^{(+)} & = u^*(k)   \qquad a_{k}^{(-)} =  i \, {\rm sign} (k) \, v(-k) \\
b_{k}^{(+)} & = - v^*(k) \qquad \hspace{-0.2cm} b_{k}^{(-)} = - i \, {\rm sign} (k) \, u(-k) \,\, .
\end{align}
\label{a_b-vs-u_v}
\end{subequations}

\vspace{-0.3cm}
\noindent
In addition, the eigenvector associated with the negative eigenvalue $E(k)=-\epsilon(k)$ can be multiplied by the overall phase factor $i \, {\rm sign} (k)$, in such a way that the unitary matrix formed by the two eigenvectors reads
\begin{equation}
D^{\dagger}(k) = 
\left( \hspace{-0.1cm} 
\begin{array}{cc}
u^*(k)    &     - v(-k)  \\
- v^*(k)  & u(-k)
\end{array}
\hspace{-0.1cm} \right) 
\label{D-matrix}
\end{equation}
with $D^{\dagger}(k) D(k) = D(k) D^{\dagger}(k) = \mathbb{1}$. 
For given $k$, $D(k)$ diagonalizes the matrix $M(k)$ of Eq.~(\ref{M-matrix}), such that
\begin{equation}
D(k) \, M(k) \, D^{\dagger}(k) = \epsilon(k)
\left( \hspace{-0.1cm} 
\begin{array}{cc}
1  &     0  \\
0  & - 1
\end{array}
\hspace{-0.1cm} \right) \, .
\label{matrix-M-diagonalized}
\end{equation}

In this way, the mean-field grand-canonical Hamiltonian of Eq.~\eqref{mean-field-total-system-Hamiltonian-1} becomes
\begin{eqnarray}
& & \hspace{-0.5cm} H^{\mathrm{mf}} - \mu N - \frac{1}{2} \sum_{k}^{BZ} \xi(k) 
 = \frac{1}{2} \sum_{k}^{BZ} \left(\begin{array}{cc} \hspace{-0.1cm} c^{\dagger}_{k} & c_{-k} \hspace{-0.1cm} \end{array}\right) M(k)
\left( \hspace{-0.1cm} 
\begin{array}{c}
c_{k}   \\
c^{\dagger}_{-k}     
\end{array}
\hspace{-0.1cm} \right)
\nonumber \\
& = & \frac{1}{2} \sum_{k}^{BZ} \, \epsilon (k) \left(\begin{array}{cc} \hspace{-0.1cm} c^{\dagger}_{k} & c_{-k} \hspace{-0.1cm} \end{array}\right) D^{\dagger}(k) 
\left( \hspace{-0.1cm} 
\begin{array}{cc}
1  &     0  \\
0  & - 1
\end{array}
\hspace{-0.1cm} \right)
D(k)
\left( \hspace{-0.1cm} 
\begin{array}{c}
c_{k}  \nonumber  \\
c^{\dagger}_{-k}     
\end{array}
\hspace{-0.1cm} \right)
\label{mean-field-total-system-Hamiltonian-2} \nonumber\\
& = & \frac{1}{2} \sum_{k}^{BZ} \, \epsilon(k) \left(\begin{array}{cc} \hspace{-0.1cm} \alpha^{\dagger}_{k} & \alpha_{-k} \hspace{-0.1cm} \end{array} \right) \!
\left( \hspace{-0.1cm} 
\begin{array}{cc}
1  &     0  \\
0  & - 1
\end{array}
\hspace{-0.1cm} \right) \!
\left( \hspace{-0.1cm} 
\begin{array}{c}
\alpha_{k}   \\
\alpha^{\dagger}_{-k}     
\end{array}
\hspace{-0.1cm} \right) \nonumber\\
&=& \sum_{k}^{BZ} \, \epsilon(k) \left( \alpha^{\dagger}_{k} \alpha_{k} - \frac{1}{2} \right) \, ,
\end{eqnarray}
where
$ \frac{1}{2} \sum_{k}^{BZ} \left( \xi(k) - \epsilon(k) \right) $ is the (grand-canonical) ground-state energy.
In the expression (\ref{mean-field-total-system-Hamiltonian-2}), we have introduced the Bogoliubov-Valatin quasi-particle operators
\begin{equation}
\left( \hspace{-0.1cm} 
\begin{array}{c}
\alpha_{k}   \\
\alpha^{\dagger}_{-k}     
\end{array}
\hspace{-0.1cm} \right)
= D(k)
\left( \hspace{-0.1cm} 
\begin{array}{c}
c_{k}   \\
c^{\dagger}_{-k}     
\end{array}
\hspace{-0.1cm} \right)
=
\left( \hspace{-0.1cm} 
\begin{array}{c}
u(k)\,c_{k} - v(k)\,c^{\dagger}_{-k}             \\
- v^*(-k)\,c_{k} + u^*(-k)\,c^{\dagger}_{-k}    
\end{array}
\hspace{-0.1cm} \right)
\label{Bogoliubov-Valatin-operators-1} \,\, ,
\end{equation}
with the anti-commutation relations
\begin{subequations}
\begin{align}
&\alpha_{k} \alpha^{\dagger}_{k'} + \alpha^{\dagger}_{k'} \alpha_{k} = \delta_{k,k'} \label{anti-commutator-relation-1} \\
&\alpha_{k} \alpha_{k'} + \alpha_{k'} \alpha_{k} = 0 \label{anti-commutator-relation-2} \,\,.
\end{align}
\end{subequations}
The expression (\ref{Bogoliubov-Valatin-operators-1}) can be inverted, to yield
\begin{equation}
\left( \hspace{-0.1cm} 
\begin{array}{c}
c_{k}   \\
c^{\dagger}_{-k}     
\end{array}
\hspace{-0.1cm} \right)
= D^{\dagger}(k) 
\left( \hspace{-0.1cm} 
\begin{array}{c}
\alpha_{k}   \\
\alpha^{\dagger}_{-k}     
\end{array}
\hspace{-0.1cm} \right)
=
\left( \hspace{-0.1cm} 
\begin{array}{c}
u^*(k)\,\alpha_{k} - v(-k)\,\alpha^{\dagger}_{-k}             \\
- v^*(k)\,\alpha_{k} + u(-k)\,\alpha^{\dagger}_{-k}    
\end{array}
\hspace{-0.1cm} \right)
\label{Bogoliubov-Valatin-operators-2} \,\, ,
\end{equation}
from which we obtain
\begin{equation}
\left\langle c_{k} c_{-k} \right\rangle = - \, u^*(k)\,v(k)
\label{average-broken-symmetry-1}
\end{equation}
where $\left\langle \cdots\right\rangle$ stands again for the average over the ground state $|\psi_{\rm GS}^{\rm mf}\rangle$ of the mean-field grand-canonical Hamiltonian \eqref{mean-field-total-system-Hamiltonian-2}, 
for which $\alpha_k|\psi_{\rm GS}^{\rm mf}\rangle=0$.
With the notation \eqref{u-notation} and~\eqref{v-notation} and the definition~\eqref{epsilon} for $\epsilon(k)$, Eq.~\eqref{average-broken-symmetry-1} becomes
\begin{equation}
\left\langle c_{k} c_{-k} \right\rangle = \frac{1}{2} \, \frac{\Delta(k)}{|\Delta(k)|} \, \sqrt{\left( 1 - \frac{\xi(k)^{2}}{\epsilon(k)^{2}} \right)} = \frac{1}{2} \, \frac{\Delta(k)}{\epsilon(k)}
\label{average-broken-symmetry-2}
\end{equation}
owing to Eq.~(\ref{k_dependent-gap-definition}), such that the gap equation~\eqref{local-gap-manipulated-1} takes the final form 
\begin{equation}
\Delta = - v_{0} \, \Delta \, \frac{1}{N} \, \sum_{k}^{BZ} \, \frac{\sin^{2}(k)}{\epsilon(k)} \,\, ,
\label{local-gap-manipulated-2}
\end{equation}
that is
\begin{equation}
\frac{1}{N} \, \sum_{k}^{BZ} \, \frac{\sin^{2}(k)}{\epsilon(k)} = -\frac{1}{v_{0}} 
\label{local-gap-manipulated-3}
\end{equation}
where $v_{0} < 0$ from the definition~\eqref{v_0-definition}.

\subsection{Role of the particle-hole symmetry}
\label{sec:particle-hole-symmetry}
Strictly in one spatial dimension, the von Neumann-Wigner theorem \cite{vN_W-1929,LL-1991} forbids the crossings of different bands.
In the present case of a fermionic p-wave fermionic superfluid, however, the occurrence of an additional (particle-hole) symmetry besides the translational symmetry allows the two branches $\pm \epsilon(k)$ 
to touch each other under special circumstances, as it was already pointed out in Ref.~\cite{Neupert-2020}.

To show this, it is relevant to consider the unitary operator
\begin{equation}
\mathcal{O}_{\mathrm{ph}} = 
\left( \hspace{-0.1cm} 
\begin{array}{cc}
0  &   1  \\
1  &   0
\end{array}
\hspace{-0.1cm} \right) \, ,
\label{particle-hole-operator}
\end{equation}
which exchanges the hole ($c_{k}$) and particle ($c^{\dagger}_{-k}$) components of the column vector
$ \left( \hspace{-0.1cm} 
\begin{array}{c}
c_{k}   \\
c^{\dagger}_{-k}     
\end{array}
\hspace{-0.1cm} \right) $
and transforms the matrix $M(k)$ of Eq.~\eqref{M-matrix} [where $\Delta^{*}(k) = - \Delta(k)$ owing to Eq.~\eqref{k_dependent-gap-definition}] in the following way:
\begin{eqnarray}
\mathcal{O}_{\mathrm{ph}} \, M(k) \, \mathcal{O}_{\mathrm{ph}} & = &
\left( \hspace{-0.1cm} 
\begin{array}{cc}
0  &   1  \\
1  &   0
\end{array}
\hspace{-0.1cm} \right) \!
\left( \hspace{-0.1cm} 
\begin{array}{cc}
\xi(k)             &     \Delta(k)  \\
- \Delta(k)    &       - \xi(k)
\end{array}
\hspace{-0.1cm} \right) \!
\left( \hspace{-0.1cm} 
\begin{array}{cc}
0  &   1  \\
1  &   0
\end{array}
\hspace{-0.1cm} \right) 
\nonumber \\
& = & - \, 
\left( \hspace{-0.1cm} 
\begin{array}{cc}
\xi(k)             &     \Delta(k)  \\
- \Delta(k)    &       - \xi(k)
\end{array}
\hspace{-0.1cm} \right) = -M(k) \,\, .
\label{particle-hole-transformation-M}
\end{eqnarray}
Accordingly, the operator $\mathcal{O}_{\mathrm{ph}}$ commutes with $M(k)$ only when $M(k)$ is the ``null'' matrix.
This occurs at the boundaries $k = (\pi,-\pi)$ of the BZ where $\Delta(k=\pm \pi) = 0$, and when $\xi(k=\pm \pi) = 2 t - \mu = 0$.
In this case, $\epsilon(k) = 0$ and the two eigenvalues $E(k) = \pm \, \epsilon(k)$ of Eq.~(\ref{diagonalization-M-matrix}) coincide with each other.

The particle-hole symmetry affects also the behavior of the functions $u(k)$ and $v(k)$ given by Eqs.~\eqref{eq:uvnotation01}, which enter the expressions (\ref{a_b-vs-u_v}) of the eigenvectors of the matrix $M(k)$.
For instance, close to the boundary $\pi$ of the BZ where $k = \pi + \kappa$ with $\kappa \ll 1$, one gets $\xi(k) \simeq 2 t - \mu - t \kappa^{2}$ and $|\Delta(k)|^{2} \simeq 4 \Delta^{2} \kappa^{2}$,
which yield $\epsilon(k) \simeq 2 \Delta |\kappa|$ when $2 t - \mu = 0$.
Accordingly,
\begin{equation}
\frac{\xi(k)}{\epsilon(k)} \simeq - t \, \frac{\kappa^{2}}{2 \Delta |\kappa|} = - \frac{t}{2 \Delta} \, |\kappa| \rightarrow 0 
\end{equation}
for $|\kappa|  \rightarrow 0$, such that $|u(k)|$ from Eq.~(\ref{u-notation}) equals $|v(k)|$ from Eq.~(\ref{v-notation}).

\section{Singe-particle wave functions}
\label{sec:wave-functions}
It is instructive to study in detail the behavior of the eigenvalues and eigenvectors of the $k$-dependent mean-field Hamiltonian matrix (\ref{M-matrix}) throughout the BZ, and determine how these quantities evolve upon varying the chemical potential $\mu$.
This behavior, in turn, affects the spatial properties of the lattice functions relevant to the fermionic p-wave fermionic superfluid of interest.
Here, we shall consider this behavior at the formal level, and later present the related numerical analysis.

\begin{figure}[t]
\includegraphics[width=8.0cm,angle=0]{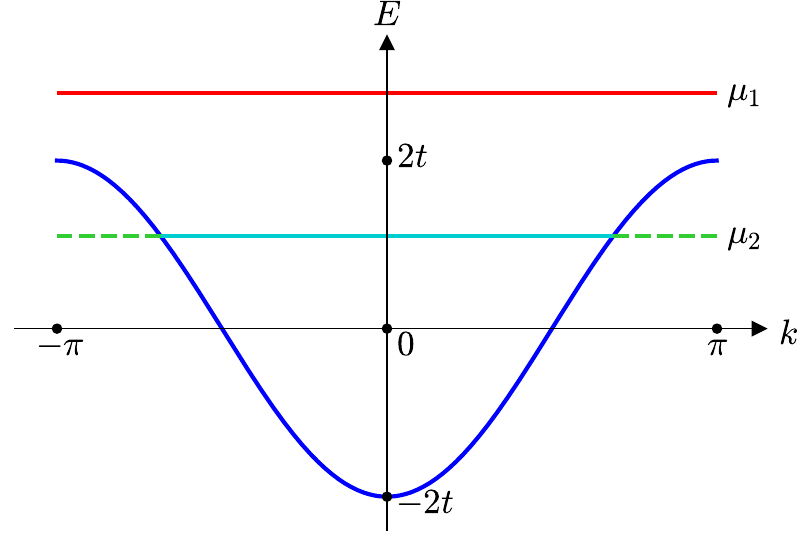}
\caption{The non-interacting dispersion relation $\lambda(k) = - 2 t \cos (k)$ is plotted in the BZ $- \pi \le k \le + \pi$. 
              Two different values of the chemical potential are considered, such that $\mu_{1} > 2t$ and $\mu_{2} < 2t$.
              In the second case, there is a region of the BZ (dashed segment) where $\xi(k) = \lambda(k) - \mu_{2} > 0$. }            
\label{Figure-1}
\end{figure} 

The sign of the non-interacting dispersion relation $\xi(k) = - 2  t \cos (k) - \mu$ (with $t>0$) plays a key role in the analysis of the eigenvalues and eigenvectors.
Two cases are possible:

\vspace{0.1cm}
\noindent
(i) $\xi(k) < 0$ for all $k$, like when $\mu = \mu_{1}$ in Fig.~\ref{Figure-1};

\vspace{0.1cm}
\noindent
(ii) there is a region in the interval $- \pi \le k \le + \pi$ where $\xi(k) >0$, like when $\mu = \mu_{2}$ in Fig.~\ref{Figure-1}.

\vspace{0.1cm}
\noindent
In both cases, since $\Delta(k=0)=0$, the ratio $\xi(k=0)/\epsilon(k=0)$ equals $\xi(k=0)/|\xi(k=0)| = {\rm sign} [\xi(k=0)]$, which is negative insofar as $\mu > - 2t$ down to the bottom of the band.

With this provision, for the present purposes it is convenient to choose the overall phases of the eigenvectors such that 
${\rm sign} (k) \sqrt{\left[ 1 + \xi(k)/\epsilon(k) \right]/2}$ is an \emph{odd\/} function of $k$ while
$\sqrt{\left[ 1 - \xi(k)/\epsilon(k) \right]/2}$ is an \emph{even\/} function of $k$, since in this way both expressions are smooth functions of $k$ about $k=0$.
Accordingly, we take again the eigenvectors of the form~\eqref{a_b-vs-u_v}, but now with the provision that 
the components of the ($+$) eigenvector are multiplied by ${\rm sign} (k)$ and
the components of the ($-$) eigenvector are multiplied by the imaginary unit.
We thus take the eigenvectors of $M(k)$ of the form:

\vspace{0.1cm}
\noindent
(i) for $E(k)=+\epsilon(k)$, multiply Eqs.~\eqref{eq:bandplusminusvector01} by ${\rm sign}(k)$, to obtain
\begin{equation}
\left( \hspace{-0.1cm} 
\begin{array}{c}
{\rm sign} (k) \, \sqrt{\frac{1}{2} \left( 1 + \frac{\xi(k)}{\epsilon(k)} \right)}  \\
i \, \sqrt{\frac{1}{2} \left( 1 - \frac{\xi(k)}{\epsilon(k)} \right)}
\end{array}
\hspace{-0.1cm} \right)
=
\left( \hspace{-0.1cm} 
\begin{array}{c}
{\rm sign} (k) \, |u(k)|  \\
i \, |v(k)|
\end{array}
\hspace{-0.1cm} \right)
\, ,
\label{+eigenvector-final}
\end{equation}

\vspace{0.1cm}
\noindent
(ii) for $E(k)=-\epsilon(k)$, multiply Eqs.~\eqref{eq:bandplusminusvector02} by $i$, to obtain
\begin{equation}
\left( \hspace{-0.1cm} 
\begin{array}{c}
i \, \sqrt{\frac{1}{2} \left( 1 - \frac{\xi(k)}{\epsilon(k)} \right)}  \\
{\rm sign} (k) \, \sqrt{\frac{1}{2} \left( 1 + \frac{\xi(k)}{\epsilon(k)} \right)}
\end{array}
\hspace{-0.1cm} \right)
=
\left( \hspace{-0.1cm} 
\begin{array}{c}
i \, |v(k)|  \\
{\rm sign} (k) \, |u(k)| 
\end{array}
\hspace{-0.1cm} \right)
\, ,
\label{-eigenvector-final}
\end{equation}
with the notation of Eqs.~(\ref{u-notation}) and (\ref{v-notation}). \revone{Note here that, although other choices of a specific $k$-dependent gauge for the eigenvectors are possible, they simply relocate the unavoidable discontinuities occurring in the BZ for the eigenvectors. In particular, if Eqs.~\eqref{eq:bandplusminusvector01} were not multiplied by ${\rm sign}(k)$, the jump discontinuity would be located at $k=0$ instead of $k =\pm \pi$ (cf also Fig.~\ref{Figure-3} below).}

Note that, when $\xi(k)$ changes sign, the eigenvectors (\ref{+eigenvector-final}) and (\ref{-eigenvector-final}) exchange with each other, with a jump of the phase by $- \pi/2$ for the upper component and by $+ \pi/2$ for the lower component,
irrespective of the sign of $k$.
This feature is associated with the presence of a \emph{quantum critical point\/} (QCP) which separates the trivial from the non-trivial topological phases, as discussed in detail below.

\subsection{Wannier-like lattice functions}
\label{sec:wannierikelatticefunctions}
The presence of the QCP is associated with an abrupt behavior in the $k$-dependence of the eigenvectors~\eqref{+eigenvector-final} and~\eqref{-eigenvector-final} across the BZ.
An alternative way for characterizing this behavior is to consider the lattice Fourier transform of the components of these eigenvectors.
This leads us to introduce the \emph{Wannier-like\/} lattice functions
\begin{eqnarray}
\hspace{-0.3cm} w_{n}^{(1)} \! & = & \! \frac{1}{N} \! \sum_{k}^{BZ} e^{i n k} {\rm sign} (k) |u(k)| = i \!\! \int_{0}^{\pi} \!\! \frac{d k}{\pi} \sin (kn)\,|u(k)| 
\label{Wannier-like-function-1}  \,\,\\
\hspace{-0.3cm}  w_{n}^{(2)} \! & = & \! \frac{1}{N} \! \sum_{k}^{BZ} e^{i n k} i |v(k)| = i \!\! \int_{0}^{\pi}\frac{d k}{\pi} \cos (kn)\,|v(k)| \,\, ,
\label{Wannier-like-function-2} 
\end{eqnarray}
associated with the components of the eigenvectors (\ref{+eigenvector-final}) and (\ref{-eigenvector-final}).
In the following, it is convenient to the absorb the imaginary unit on the left-hand side of Eqs.~(\ref{Wannier-like-function-1}) and (\ref{Wannier-like-function-2}), by redefining $w_{n}^{(j)} \, \rightarrow \, - i \, w_{n}^{(j)}$ with $j=1,2$.

\subsection{Pair wave function and correlation function}
\label{sec:pairwavecorrelationfunction}
Together with the spatial behavior of the single-particle wave functions, it is relevant to consider also the spatial behavior of the \emph{pair wave function\/} $g(n)$.
This is obtained by recalling the expression of the BCS ground state \cite{BCS-1957}
\begin{eqnarray}
| \Phi_{\mathrm{BCS}} \rangle & = & \prod_{k \ge 0}^{BZ} \left( u(k) + v(k) \, c^{\dagger}_{k} c^{\dagger}_{-k} \right) \! | 0 \rangle
\nonumber \\
& = & \frac{1}{\mathcal{N}} \, \prod_{k \ge 0}^{BZ}  \left( 1 + \frac{v(k)}{u(k)} \, c^{\dagger}_{k} c^{\dagger}_{-k} \right) \! | 0 \rangle
 \nonumber \\
& = & \frac{1}{\mathcal{N}} \, \prod_{k \ge 0}^{BZ} \, e^{ \frac{v(k)}{u(k)} \, c^{\dagger}_{k} c^{\dagger}_{-k} } \, | 0 \rangle
=  \frac{1}{\mathcal{N}} \, e^{ \sum_{k \ge 0}^{BZ} \, \frac{v(k)}{u(k)} \, c^{\dagger}_{k} c^{\dagger}_{-k} } \, | 0 \rangle \,\, ,
\nonumber\\
\label{BCS-ground-state}
\end{eqnarray}
where $| 0 \rangle$ is the fermionic vacuum, $\mathcal{N} = \left( \prod_{k \ge 0}^{BZ} u(k) \right)^{-1}$ a normalization factor, and $u(k)$ and $v(k)$ are the same functions of Eqs.~(\ref{eq:uvnotation01}).
This can be verified by calculating the average value of the operator $c_{k} c_{-k}$ over the BCS ground state (\ref{BCS-ground-state}), recovering in this way the expression (\ref{average-broken-symmetry-1}).
With the lattice Fourier transform of Eq.~\eqref{lattice-Fourier-transform}, the expression in the exponent on the right-hand side of Eq.~\eqref{BCS-ground-state} becomes
\begin{equation}
\sum_{k \ge 0}^{BZ} \, \frac{v(k)}{u(k)} \, c^{\dagger}_{k} c^{\dagger}_{-k} = \frac{1}{2} \sum_{k}^{BZ} \, \frac{v(k)}{u(k)} \, c^{\dagger}_{k} c^{\dagger}_{-k} 
= \sum_{n n'} \, g(n-n') \, c^{\dagger}_{n} c^{\dagger}_{n'} \,,
\label{Fourier-transform-pair-wave-function}
\end{equation}
where
\begin{equation}
g(n) \equiv \frac{1}{2N} \, \sum_{k}^{BZ} e^{i k n} \, \frac{v(k)}{u(k)} = - \! \int_{0}^{\pi} \! \frac{d k}{2 \pi} \, \sin (k n) \, \frac{|v(k)|}{|u(k)|} 
\label{definition-g}
\end{equation}
because the ratio $v(k)/u(k)$ is an odd function of $k$ since $u(k)$ is even and $v(k)$ is odd [cf. Eqs.~\eqref{eq:uvnotation01}].
Recall also that the BCS ground state (\ref{BCS-ground-state}) is annihilated by the Bogoliubov-Valatin quasi-particle operator $\alpha_{k}$ of Eq.~(\ref{Bogoliubov-Valatin-operators-1}), namely,
\begin{eqnarray}
& & \hspace{-0.5cm} \alpha \, | \Phi_{\mathrm{BCS}} \rangle =  \left( u(k) c_{k} - v(k) c^{\dagger}_{-k}  \right) | \Phi_{\mathrm{BCS}} \rangle
\label{alpha-annihilates-Phi_BCS} 
\nonumber \\
& \hspace{-0.5cm} = & u(k) v(k) \!\! \prod_{k' \ne k \ge 0}^{BZ} \! \left( u(k') + v(k') \, c^{\dagger}_{k'} c^{\dagger}_{-k'} \right) c_{k} c^{\dagger}_{k} c^{\dagger}_{-k} | 0 \rangle 
\nonumber \\
& \hspace{-0.5cm} - & u(k) v(k) \!\!  \prod_{k' \ne k \ge 0}^{BZ} \! \left( u(k') + v(k') \, c^{\dagger}_{k'} c^{\dagger}_{-k'} \right) c^{\dagger}_{-k} | 0 \rangle = 0 \,\, ,
\label{annihilation-ground-state}
\end{eqnarray}
where we have again emphasized that $\prod_{k'}$ covers only half of the BZ.

\vspace{0.1cm}
It is further of interest to consider the correlation function
\begin{equation}
F(x,x') = \left\langle \Psi(x) \, \Psi(x') \right\rangle \,\, ,
\label{correlation-function-1}
\end{equation}
which reads within the one-band approximation (\ref{one-band-approximation}) 
\begin{equation}
F(x,x') = \sum_{n n'} \, \phi(x - n) \, \left\langle c_{n} \, c_{n'} \right\rangle \, \phi(x' - n') \,\, .
\label{correlation-function-2}
\end{equation}
By approximating $\int \!\! d x \, \phi(x - m) \, \phi(x - n) \simeq \delta_{n,m}$, from the expression (\ref{correlation-function-2}) we can extract the quantity
\begin{equation}
\int \!\! d x dx' \phi(x - m) \, F(x,x') \, \phi(x' - m') \simeq \left\langle c_{m} \, c_{m'} \right\rangle \,\, ,
\label{correlation-function-3}
\end{equation}
which generalizes to arbitrary $m'$ (beyond the nearest neighbor to $m$) the quantity $\left\langle c_{m} \, c_{m+1} \right\rangle$ entering the local gap (\ref{local-gap-definition}). 
Exploiting translational invariance and the lattice Fourier transform (\ref{lattice-Fourier-transform}) as we did in Eq.~(\ref{local-gap-manipulated-1}), 
the right-hand side of Eq.~(\ref{correlation-function-3}) becomes
\begin{eqnarray}
& & \hspace{-0.4cm} f(n) = \frac{1}{N} \sum_{m} \, \langle c_{m} c_{m+n} \rangle  = \frac{1}{N} \sum_{k}^{BZ} \, e^{-ikn} \, \langle c_{k} c_{-k} \rangle 
\label{definition-f} \nonumber \\
& \hspace{-0.4cm} = & - \frac{1}{N} \sum_{k}^{BZ} \! e^{-ikn} \, u(k)^{*} v(k) = - \!\! \int_{0}^{\pi} \!\! \frac{d k}{\pi} \! \sin (k n) |u(k)| \, |v(k)| \,\, ,\nonumber\\
\label{correlation-function-4}
\end{eqnarray}
where the result~(\ref{average-broken-symmetry-1}) has been utilized.

The quantities $g(n)$ from Eq.~(\ref{definition-g}) and $f(n)$ from Eq.~(\ref{definition-f}) both vanish for $n = 0$ due to Pauli principle, but a have different spatial behavior for $n \ge 1$.
This can be explicitly verified by simple analytic examples, where $g(n)$ and $f(n)$ are obtained exactly by selecting appropriate forms of the functions $u(k)$ and $v(k)$ as follows:

\subsubsection*{Example 1: $\mu=0$}
In this case, the eigenvalue equation (\ref{diagonalization-M-matrix}) can be solved analytically, to obtain $u(k) = \sin \! \left(\frac{k}{2}\right)$ and $v(k) = i \, \cos \! \left(\frac{k}{2}\right)$ for $k > 0$.
Accordingly, we obtain 
\begin{equation}
- 2 \pi g(n) = \! \int_{0}^{\pi} \!\! d k \, \sin (k n) \, \frac{\cos \! \left(\frac{k}{2}\right)}{\sin \! \left(\frac{k}{2}\right)} = \pi \qquad (n \ge 1)
\label{g_n-model-1}
\end{equation}
where the result \#~3.612.7 at page~366 of Ref.~\cite{Gradshteyn-Ryzhik-1980} has been utilized, in agreement with the general result (\ref{general-asymptotic-result-g}) for $\mu \le 1$ discussed later on.

In a similar manner, we obtain
\begin{eqnarray}
- \pi f(n) &=& \! \int_{0}^{\pi} \!\! d k \, \sin (k n) \, \sin \! \left(\frac{k}{2}\right) \, \cos \! \left(\frac{k}{2}\right) \nonumber\\
&=&
\left\{ 
\begin{array}{ll}
\frac{\pi}{4}  &  \mathrm{for} \, n = 1 \\
0                  &  \mathrm{for} \, n > 1
\end{array}
\right. \, ,
\label{f_n-model-1}
\end{eqnarray}
where the results \#~4.3.140 and \#~4.3.141 at page~78 of Ref.~\cite{Abramowitz-Stegun-1972} have instead been utilized.
Note the different ``asymptotic'' behavior between the results (\ref{g_n-model-1}) and (\ref{f_n-model-1}).

\subsubsection*{Example 2: $\mu\gg1$}
In this case, we consider the approximate expressions $u(k) = \gamma \sin (k)$ and $v(k) = i \left( 1 - \frac{\gamma^{2}}{2} \sin^{2}(k) \right)$ with $\gamma \ll 1$
(which reduces to the non-interacting case in the limit $\gamma \, \rightarrow \, 0$), such that
\begin{eqnarray}
- 2 \pi g(n) & = & \! \int_{0}^{\pi} \!\! d k \, \sin (k n) \, \frac{\left( 1 - \frac{\gamma^{2}}{2} \sin^{2}(k) \right)}{\gamma \sin (k)} 
\nonumber \\
& = & \frac{1}{\gamma} \! \int_{0}^{\pi} \!\! d k \, \frac{\sin (k n)}{\sin (k)} - \frac{\gamma}{2} \! \int_{0}^{\pi} \!\! d k \, \sin (k n) \, \sin (k)
\nonumber \\
& = & - \frac{\pi \, \gamma}{4} \, \delta_{n,1} + \frac{1}{\gamma} 
\left\{ 
\begin{array}{ll}
\pi  &  \mathrm{for} \, n \, \mathrm{odd}  \\
0    &  \mathrm{for} \, n \, \mathrm{even}
\end{array}
\right. \,\, ,
\label{g_n-model-2}
\end{eqnarray}
where the result~\# 3.612.2 at page~366 of Ref.~\cite{Gradshteyn-Ryzhik-1980} has been utilized. 
In addition, we obtain
\begin{eqnarray}
\hspace{-0.5cm} - \pi f(n) & = & \! \int_{0}^{\pi} \!\! d k \, \sin (k n) \,\, \gamma \sin (k) \left( 1 - \frac{\gamma^{2}}{2} \sin^{2}(k) \right) 
\nonumber \\
& \simeq & \gamma \! \int_{0}^{\pi} \!\! d k \, \sin (k n) \, \sin (k) =
\left\{ 
\begin{array}{ll}
\frac{\gamma \, \pi}{2}  &  \mathrm{for} \, n = 1 \\
0                                   &  \mathrm{for} \, n > 1 
\end{array}
\right. .
\label{f_n-model-2}
\end{eqnarray}

Note that $f(n)$ is of short range in both cases, while $g(n)$ extends up to infinity although in different ways in the two cases.

These analytic results are in line with the numerical results obtained in the rest of this paper, when solving numerically for the eigenvectors of the matrix (\ref{M-matrix})
(cf. Fig.~\ref{Figure-5} below).

\section{Numerical results for the p-wave fermionic superfluid}
\label{sec:numerics-superfluid-phase}
We now report the numerical results for the quantities considered in Eqs.~\eqref{Wannier-like-function-1}, \eqref{Wannier-like-function-2}, \eqref{definition-g}, and~\eqref{definition-f}. 
Special emphasis will be given to the emergence of discontinuities in the BZ when passing from the trivial to the non-trivial topological phases,
as well as to the related spatial falloff of the associated Wannier-like functions and correlation functions.
For definiteness, in the following we shall take $2 t = 2 \Delta = 1$, such that $\xi(k) \rightarrow -\cos (k) -\mu$ and $\Delta(k) \rightarrow - i \, \sin (k)$.
In addition, we will dispose of the imaginary unit in Eqs.~(\ref{Wannier-like-function-1}) and (\ref{Wannier-like-function-2}), and of the minus sign in Eqs.~(\ref{definition-g}) and (\ref{correlation-function-4}),
by defining $|w_{n}^{(1)}| = - i w_{n}^{(1)}$, $|w_{n}^{(2)}| = - i w_{n}^{(2)}$, $|g(n)| = - g(n)$, and $|f(n)| = - f(n)$.

\subsection{Evolution of the eigenvectors in the Brillouin zone}
\label{eigenvectorbrillouin}
\begin{figure*}[t]
\begin{center}
\includegraphics[width=17.9cm,angle=0]{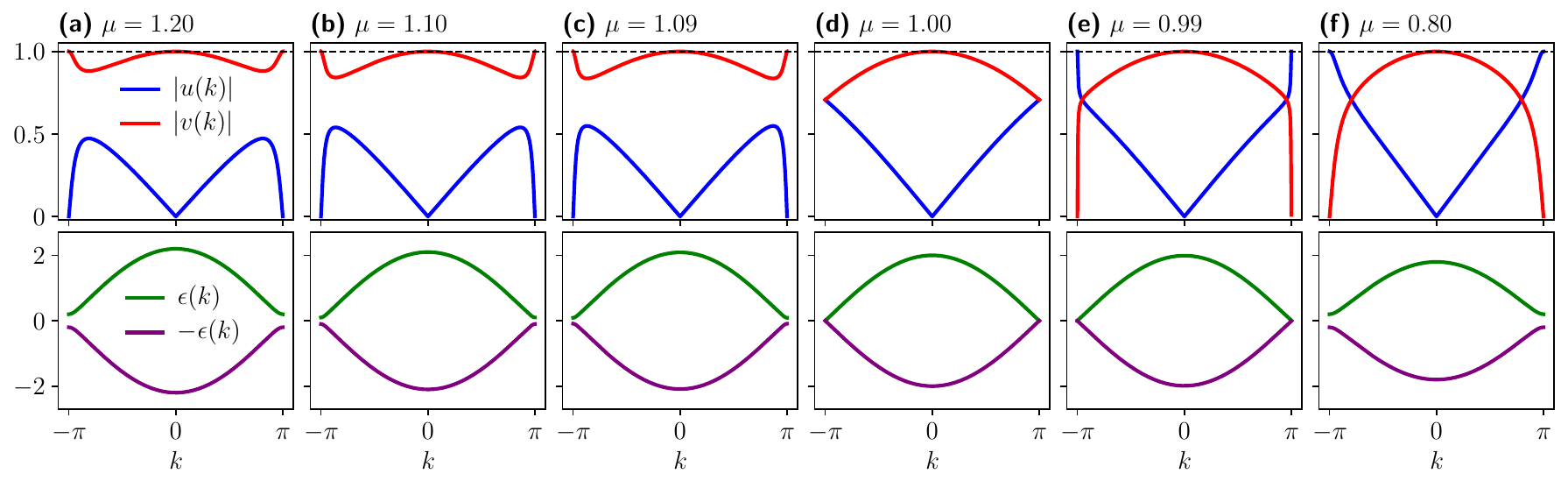}
\caption{The functions $|u(k)|$ (blue lines) and $|v(k)|$ (red lines) from Eqs.~(\ref{u-notation}) and (\ref{v-notation}) are plotted in the top panels over the whole BZ, for several values of $\mu$ across the QCP at $\mu=1$
             (with the horizontal dashed line marking the unit value).   
              The corresponding eigenvalues $E(k) = \pm \epsilon(k)$ from Eq.~(\ref{epsilon}) are plotted in the bottom panels. 
              In (the pairs of) panels from \textbf{(a)} to \textbf{(c)} corresponding to the trivial phase with $\mu>1$, $|u(k)|$ and $|v(k)|$ do not cross each other and the eigenvalues $\pm\epsilon(k)$ have a gap 
              $\Delta E = 2 \, \epsilon(k=\pm \pi)$. 
              In (the pairs of) panels \textbf{(d)}, where the QCP occurs at $\mu=1$, $|u(k)|$ and $|v(k)|$ touch each other at $k=\pm\pi$ and the energy gap closes. 
              In this case, $|u(k=\pm\pi)|=|v(k=\pm\pi)| = 1/\sqrt{2}$ are both nonzero. 
              In (the pairs of) panels from \textbf{(e)} to \textbf{(f)} corresponding to the non-trivial phase with $\mu < 1$ past the QCP, $|u(k)|$ and $|v(k)|$ cross each other at some value of $k$ inside the BZ and the energy gap opens again.}
\label{Figure-2}
\end{center} 
\end{figure*}

Figure~\ref{Figure-2} shows the functions $|u(k)|$ and $|v(k)|$ entering the expressions~(\ref{+eigenvector-final}) and (\ref{-eigenvector-final}) of the eigenvectors 
along the whole BZ, together with the corresponding eigenvalues $E(k) = \pm \epsilon(k)$ with $\epsilon(k)$ given by Eq.~(\ref{epsilon}).
Several cases for both $\mu>1$ and $\mu\leq1$ are considered, with the QCP occurring at $\mu = 1$.
It turns out that for $\mu>1$ the system is in the \emph{trivial phase} where $|u(k)|$ and $|v(k)|$ never cross each other, while for $\mu \le 1$ the system is in the \emph{non-trivial phase} where $|u(k)|$ and $|v(k)|$ do cross each other at an intermediate value of $k$ inside the Brillouin zone.
Pictorially, one could say that the eigenvectors components pass from being ``unlocked'' in the trivial phase (cf. panels \textbf{(a)}-\textbf{(c)} of Fig.~\ref{Figure-2}) to be ``locked'' in the non-trivial phase (cf. panels \textbf{(d)}-\textbf{(f)} of Fig.~\ref{Figure-2}),
where this drastic change is at the essence of the topological transition across the QCP.

Specifically, in the trivial phase Fig.~\ref{Figure-2} shows that, for increasing values of $\mu>1$, $|u(k)|$ progressively approaches zero while $|v(k)|$ approaches unity for all $k$, thereby asymptotically recovering the non-interacting case with $\Delta \rightarrow 0$ and $\xi(k) <0$ for all $k$. 
Furthermore, $|u(k=0)| = |u(k=\pi)| = 0$ while $|v(k=0)| = |v(k=\pi)| = 1$, such that the eigenvectors~\eqref{+eigenvector-final} and~\eqref{-eigenvector-final} can be smoothly continued from positive
to negative $k$, with no discontinuity occurring not only at $k=0$ but also at $k = \pi$.
This situation, shown in panels from \textbf{(a)} to \textbf{(c)} of Fig.~\ref{Figure-2}, will imply that the Wannier-like lattice functions (\ref{Wannier-like-function-1}) and (\ref{Wannier-like-function-2}) have a \emph{short-range} spatial behavior
(as shown in Fig.~\ref{Figure-4} below), with a similar behavior occurring for the pair wave function and the correlation function (as shown in Figs.~\ref{Figure-5} and~\ref{Figure-6} below).

In the non-trivial phase, on the other hand, there is no way of avoiding either a jump discontinuity or the presence of a cusp in the eigenvectors, once they are extended to the whole BZ.
This situation, shown in panels from \textbf{(d)} to \textbf{(f)} of Fig.~\ref{Figure-2}, will imply a \emph{long-range} spatial behavior for the Wannier-like lattice functions, while both the pair wave function and the correlation function maintain a short-range spatial behavior for $\mu < 1$ (as shown in Figs.~\ref{Figure-5} and~\ref{Figure-6} below).

\begin{figure}[b]
\begin{center}
\includegraphics[width=8.5cm,angle=0]{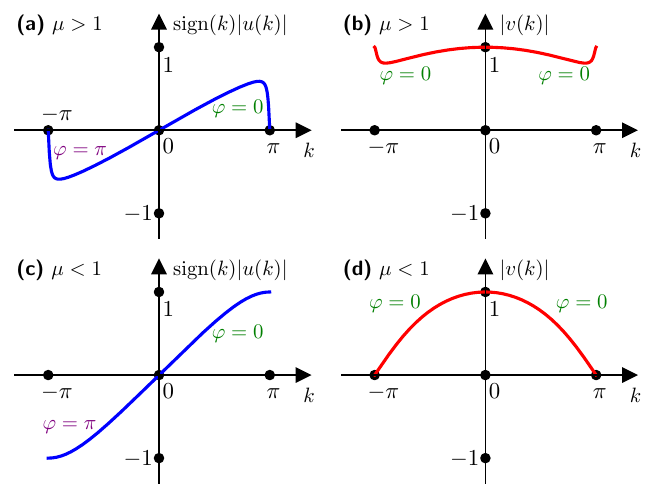}
\caption{The eigenvectors (\ref{+eigenvector-final}) for the positive eigenvalue $+ \epsilon(k)$ are shown over the whole BZ, for two typical cases corresponding to \textbf{(a)}-\textbf{(b)} $\mu > 1$ and \textbf{(c)}-\textbf{(d)} $\mu < 1$. 
             The corresponding values of the overall phase $\varphi$ of the wave-function components are also shown in both halves of the BZ.
             }
\label{Figure-3}
\end{center} 
\end{figure} 

In addition, Fig.~\ref{Figure-3} reports the overall phase $\varphi$ of the eigenvector components in both halves of the BZ for $\mu >1$ (trivial phase) and $\mu<1$ (non-trivial phase), where,
in line with the expression (\ref{+eigenvector-final}) of the eigenvector with positive eigenvalue, $|u(k)|$ is multiplied by sign$(k)$.
While in all cases no discontinuity in the value of $\varphi$ occurs for $|v(k)|$ (cf. panels \textbf{(b)} and \textbf{(d)} of Fig.~\ref{Figure-3}), for sign$(k) |u(k)|$ in both cases $\varphi$ jumps by $\pi$ at $k=\pi$ and by $-\pi$ at $k=0$ (cf. panels \textbf{(a)} and \textbf{(c)} of Fig.~\ref{Figure-3}), 
such that $d\varphi/dk = - \pi\,\delta(k) + \pi\,\delta(k-\pi)$. 
However, while in the trivial phase of panel \textbf{(a)} $|u(k=0)|^{2}=|u(k=\pi)|^{2}=0$ such that these phase jumps do not contribute to the Berry phase, in the non-trivial phase of panel \textbf{(c)} $|u(k=0)|^{2}=0$ and $|u(k=\pi)|^{2}=1$, 
yielding the value $-\pi$ for the Berry phase~\cite{Neupert-2020}.

\subsection{Lattice-site dependence of the Wannier-like functions}
\label{sec:numerics2}
The change in the topological properties of the eigenvectors leads to a corresponding abrupt modification of the site dependence of the Wannier-like functions~(\ref{Wannier-like-function-1}) and (\ref{Wannier-like-function-2}), from short to long range.
Here, we quantify this spatial behavior in detail, by evaluating numerically the expressions~(\ref{Wannier-like-function-1}) and (\ref{Wannier-like-function-2}).

\begin{figure*}[t]
\begin{center}
\includegraphics[width=17.2cm,angle=0]{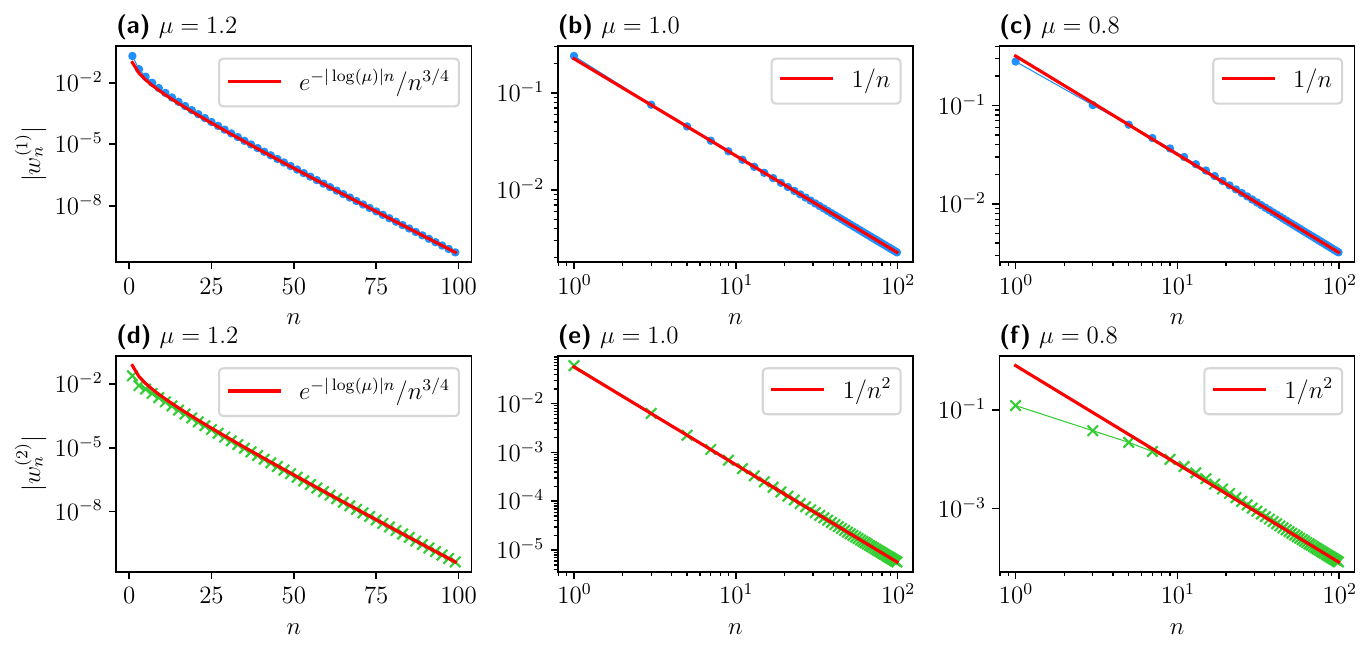}
\caption{Dependence on the lattice sites $n$ of the magnitude of the Wannier-like functions~(\ref{Wannier-like-function-1}) (panels from \textbf{(a)} to \textbf{(c)}) and (\ref{Wannier-like-function-2}) (panels from \textbf{(d)} to \textbf{(f)}).  
              Data from the numerical evaluation of the integrals in Eqs.~(\ref{Wannier-like-function-1}) and (\ref{Wannier-like-function-2}) are shown as blue dots and green crosses, respectively, 
              while the expected asymptotic behavior obtained analytically is shown by a red solid line in all cases. 
              Pairs of panels correspond to three values of $\mu$ across the QCP at $\mu=1$.
              To highlight the exponential decay for large $n$ of the Wannier-like functions when $\mu>1$, in panels \textbf{(a)} and \textbf{(d)} a log-linear scale is used. 
              To highlight instead the power-law decay for large $n$ of the Wannier-like functions when $\mu\leq1$, in the remaining panels from \textbf{(b)} to \textbf{(f)} a log-log scale is used. 
              In all cases, the analytic behavior for large-$n$ (obtained as explained in the text) is reported in the boxes.}
\label{Figure-4}
\end{center} 
\end{figure*}

Figure~\ref{Figure-4} reports the results of this numerical analysis.
Specifically, the upper panels from \textbf{(a)} to \textbf{(c)} show $|w^{(1)}_n|$, while the lower panels from \textbf{(d)} to \textbf{(f)} show $|w^{(2)}_n|$, for three different values of $\mu$ across the QCP at $\mu=1$. 
Data from the numerical evaluation of the integrals in Eqs.~(\ref{Wannier-like-function-1}) and (\ref{Wannier-like-function-2}) are indicated by blue dots for $|w^{(1)}_n|$ and green crosses for $|w^{(2)}_n|$, 
while the expected asymptotic behavior for large $n$ (reported in the box in each panel) is shown in all cases by a red solid line. 

In particular, in the trivial phase for $\mu>1$, panels \textbf{(a)} and \textbf{(d)} show that both Wannier-like functions falloff exponentially for large $n$, thus demonstrating a short-range behavior 
when the functions $|u(k)|$ and $|v(k)|$ can be smoothly connected from $k=-\pi$ to $k=\pi$, as mentioned before. 
On the other hand, in the non-trivial phase for $\mu\leq1$, panels \textbf{(b)} and \textbf{(c)} for $|w^{(1)}_n|$ and panels \textbf{(e)} and \textbf{(f)} for $|w^{(2)}_n|$ show that the large-$n$ falloff of the Wannier-like functions 
follows a power-law behavior, as a consequence of the fact that the functions $|u(k)|$ and $|v(k)|$ cannot be smoothly connected from $k=-\pi$ to $k=\pi$.
In all panels of Fig.~\ref{Figure-4}, the asymptotic (either exponential or power-law) behavior of the Wannier-like functions for large $n$ is indicated by red lines.

These asymptotic behaviors can be obtained analytically from the methods of Refs.~\cite{Lighthill-1958,Boyd-2009}, by analyzing the most relevant discontinuity of the functions ${\rm sign}(k)|u(k)|$ and $|v(k)|$ over the whole BZ. 
Specifically, there are at most \emph{three discontinuities\/} occurring in the functions ${\rm sign}(k)|u(k)|$ and $|v(k)|$ for different values of $\mu$.
With reference to Fig.~\ref{Figure-2}, they can be classified as follows:

\vspace{0.1cm}
\noindent
(i) The poles $z_s$ of order one of $1/\epsilon^2(z)$, which arise from the zeros of $\epsilon^2(z)=1+\mu^2+2\mu\cos(z)$ in Eq.~\eqref{epsilon} (with $2 t = 2 \Delta = 1$) when $k \rightarrow z$ is extended over the complex $z$-plane, 
and are given by $z_s=(2s+1)\pi\pm i\log(\mu)$ ($s$ integer).
In this way, the function $1/\sqrt{\epsilon(z)}$ entering in the definition of $|u(z)|$ and $|v(z)|$ in Eqs.~\eqref{u-notation} and~\eqref{v-notation} has a non-analytic behavior of the form $1/{(z-z_s)}^{1/4}$, yielding
an asymptotic behavior of the form $e^{-|\log(\mu)|n}/n^{3/4}$ for the Fourier transforms.
It should be mentioned that a branch-point singularity of the form $1/{(z-z_s)}^{1/4}$ was anticipated in Ref.~\cite{Kohn-1959} while discussing the analytic properties of Bloch wave and Wannier functions for non-interacting fermions in one
dimension, for which the energy bands never cross each other for real $k$. 

\vspace{0.05cm}
\noindent
(ii) The sign discontinuity of ${\rm sign}(k)|u(k)|$ occurring in the non-trivial phase at $k=\pm\pi$, yielding a $1/n$ power-law asymptotic behavior for the Fourier transform.

\vspace{0.05cm}
\noindent
(iii) The sign discontinuity of the first derivative of $|v(k)|$ in the non-trivial phase at $k=\pm\pi$, yielding a $1/n^2$ power-law asymptotic behavior for the Fourier transform.

\vspace{0.05cm}
According to the above analysis, in the trivial phase, when only the non-analytic behavior due to the poles of $1/\sqrt{\epsilon(z)}$ is present, the asymptotic behavior of both Wannier-like functions is given by the exponential falloff 
$e^{-|\log(\mu)|n}/n^{3/4}$ as long as $\mu > 1$, as reported in panels \textbf{(a)} and \textbf{(d)} of Fig.~\ref{Figure-4}. 
On the other hand, in the non-trivial phase the sign discontinuity of ${\rm sign}(k)|u(k)|$ makes the power-law falloff $1/n$ to dominate over the exponential one associated with the poles of $1/\sqrt{\epsilon(z)}$, 
yielding a $1/n$ asymptotic behavior of the Wannier-like function $|w^{(1)}_n|$, as reported in panels \textbf{(b)} and $\textbf{(c)}$ of Fig.~\ref{Figure-4}. 
Similarly, the sign discontinuity of $d|v(k)|/dk$ makes the power-law falloff $1/n^2$ to dominate over the exponential one, yielding a $1/n^2$ asymptotic behavior of the Wannier-like function $|w^{(2)}_n|$, 
as reported in panels \textbf{(e)} and $\textbf{(f)}$ of Fig.~\ref{Figure-4}.

\subsection{Pair wave function and the correlation function}
\label{sec:numerics3}
\begin{figure*}[t]
\begin{center}
\includegraphics[width=17.2cm,angle=0]{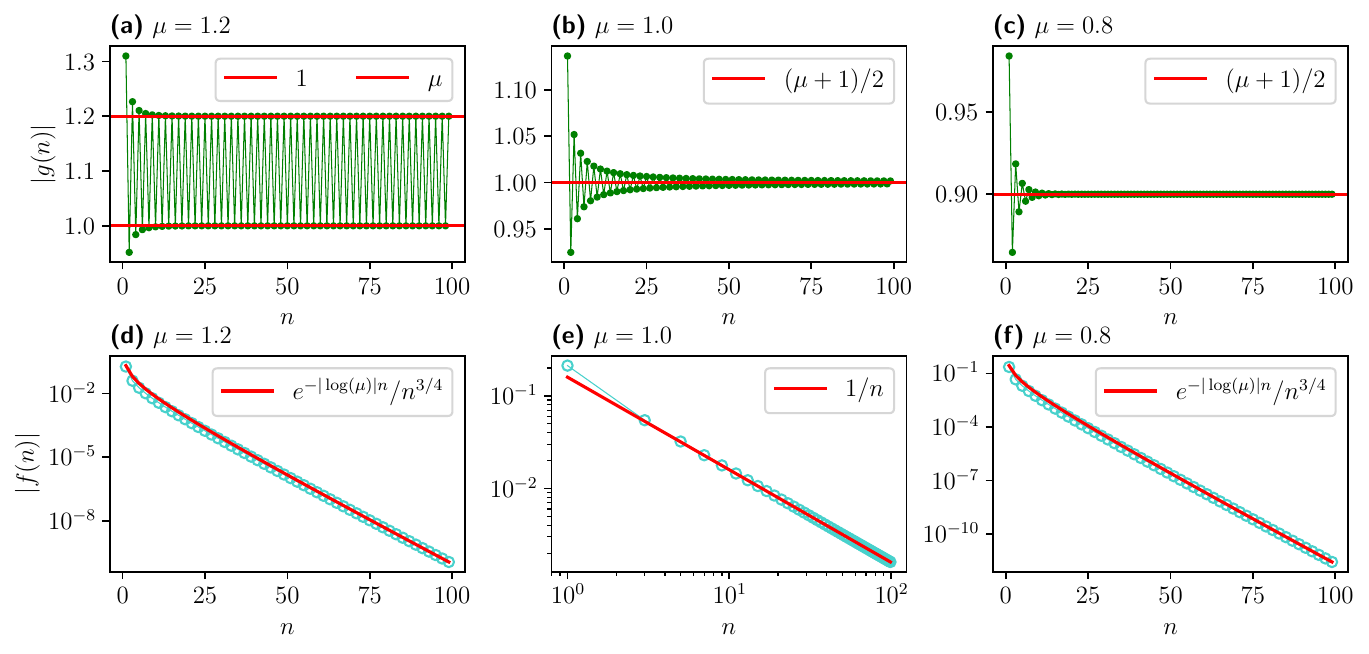}
\caption{Dependence on the lattice site $n$ of the pair wave function $|g(n)|$ from Eq.~\eqref{definition-g} (panels from \textbf{(a)} to \textbf{(c)}) and of the correlation function $|f(n)|$ from Eq.~\eqref{definition-f} (panels from \textbf{(d)} to \textbf{(f)}). 
              Data from the numerical evaluation of the integrals in Eq.~\eqref{definition-g} and~\eqref{definition-f} are shown by green dots and cyan open circles, respectively, while the expected asymptotic behavior obtained analytically is shown by red solid lines 
              in all cases.
              Like in Fig.~\ref{Figure-4}, pairs of panels correspond to three values of $\mu$ across the QCP at $\mu=1$.
              Data for $|g(n)|$ are reported in a linear scale, showing an alternating behavior for even and odd $n$ when $\mu>1$ in panel \textbf{(a)}, and the reaching of a single uniform value for $\mu\leq1$ in panels \textbf{(b)} and \textbf{(c)}. 
              Data for $|f(n)|$ are reported in a log-linear scale in panel \textbf{(d)} for $\mu>1$ and in panel \textbf{(f)} for $\mu<1$ to highlight an exponential falloff for large $n$, while data for $\mu=1$ are reported in panel \textbf{(e)} in a log-log scale to highlight 
              a power-law falloff. 
              In addition, the asymptotic falloff of $|f(n)|$ for large-$n$ is given in the boxes as explained in the text (while the asymptotic falloff of $|g(n)|$ for large-$n$ will be discussed in Fig.~\ref{Figure-6}).}
\label{Figure-5}
\end{center} 
\end{figure*}

We next consider the corresponding analysis for (the absolute value of) the pair wave function $|g(n)|$ of Eq.~\eqref{definition-g} and the correlation function $|f(n)|$ of Eq.~\eqref{definition-f}. 
Figure~\ref{Figure-5} reports the results obtained by numerical evaluation of the integrals in those expressions, with the respective large-$n$ behavior obtained by arguments similar to those considered in Fig.~\ref{Figure-4} for the Wannier-like functions. 

From panel \textbf{(a)} of Fig.~\ref{Figure-5} with $\mu > 1$, we see that for large $n$ $|g(n)|$ displays a pattern that oscillates between the values $1$ and $\mu$ for even and odd $n$, respectively, with average value $(\mu+1)/2$. 
On the other hand, from panels \textbf{(b)} and \textbf{(c)} of Fig.~\ref{Figure-5} with $\mu \le 1$, we see that $|g(n)|$ approaches the single uniform value $(\mu+1)/2$.
In all cases, the function $|g(n)|$ remains finite for $n$ extending up to infinity.
These peculiar trends are confirmed by the analytic results reported for the two simple cases with $\mu=0$ and $\mu\gg1$ discussed above.

In addition, from panels from \textbf{(d)} to \textbf{(f)} of Fig.~\ref{Figure-5}, we see that for large $n$ the function $|f(n)|$ falloffs to zero in all cases. 
In particular, $|f(n)|$ is short-ranged when $\mu\neq1$ (cf. panels \textbf{(d)} and \textbf{(f)} of Fig.~\ref{Figure-5}) with the same exponential behavior found for the Wannier-like functions in Fig.~\ref{Figure-4}, 
while $|f(n)|$ falloffs as a power law when $\mu=1$ (cf. panel \textbf{(e)} of Fig.~\ref{Figure-5}).
In all panels, the analytic behavior of $|f(n)|$ for large $n$ is shown by red solid lines, which are drawn using the same arguments as for the Wannier-like functions in Fig.~\ref{Figure-4}. 
Specifically, since $|f(n)|$ is the lattice Fourier transform of the \emph{odd} function $v(k)\,u(k)$ for $k$ extending over the whole BZ, with the functions $|u(k)|$ and $|v(k)|$ evolving across the QCP like in Fig.~\ref{Figure-2}, 
we conclude that:

\vspace{0.1cm}
\noindent
(i) When $\mu\neq1$, the product $v(k)\,u(k)$ is zero not only at $k=0$ but also at $k=\pm\pi$, because $|u(k=\pm\pi)|=0$ when $\mu>1$ and $|v(k=\pm\pi)|=0$ when $\mu<1$. 
As a consequence, the product $v(k)\,u(k)$ is smooth across the whole Brillouin zone, with no sign discontinuity. 
As discussed before, the dominant discontinuity is then that arising from the poles of $1/\sqrt{\epsilon(z)}$ in the complex plane, yielding the exponential falloff $e^{-|\log(\mu)|n}/n^{3/4}$ for large $n$.

\vspace{0.1cm}
\noindent
(ii) When $\mu=1$, the product $v(k)\,u(k)$ is nonzero at $k=\pm\pi$, because in this case $|u(k=\pm\pi)| = |v(k=\pm\pi)| = 1/\sqrt{2}$. 
As a consequence, the product $v(k)\,u(k)$ has a sign discontinuity at $k=\pm\pi$, yielding the dominant $1/n$ power-law behavior for large $n$.

The large-$n$ behavior of $|g(n)|$ (as reported in the top panels from \textbf{(a)} to \textbf{(c)} of Fig.~\ref{Figure-5}) can be obtained by the following considerations. 
When restricting to the positive half of the Brillouin zone where $k\geq0$, the integrand function $\sin(kn) |v(k)|/|u(k)|$ in Eq.~\eqref{definition-g} has removable singularities of the form $\sin(nx)/x$ at $k=0$ for all $\mu$ and at $k=\pi$ for $\mu>1$, 
where $|u(k)|=0$. 
This can be seen by expanding the function $\mathcal{G}(k) \equiv |v(k)|/|u(k)|$ for $k\rightarrow0^+$ and $k\rightarrow\pi^-$. 
For the leading behavior, in the first ($k\rightarrow0^+$) case we obtain
\begin{equation}
\mathcal{G}(k) \simeq \frac{2(1+\mu)}{k} \qquad (\mbox{all $\mu$}) \,\, ,
\label{eq:smallkexpansion1}
\end{equation}
while in the second ($k\rightarrow\pi^-$) case we obtain
\begin{equation}
\mathcal{G}(k) \simeq \left\{\begin{array}{ll}
\frac{2(\mu-1)}{\pi-k} & \quad (\mu>1) \vspace{0.2cm}\\
1+\frac{\pi-k}{2} & \quad (\mu=1) \vspace{0.2cm}\\
\frac{\pi-k}{2(1-\mu)} & \quad (\mu<1)
\end{array}\right. \,\, .
\label{eq:smallkexpansion2}
\end{equation}
In the limit $n\gg1$, the main contribution to the integral $2 \pi|g(n)| = \int_{0}^{\pi}dk\,\sin(nk) \, \mathcal{G}(k)=n^{-1}\int_{0}^{n\pi}dp\,\sin(p) \, \mathcal{G}(p/n)$ (with $p=nk$) originates from the neighborhoods of these values of $k$ 
where the removable singularity occurs. 
We then obtain in the non-trivial ($\mu \le 1$) and trivial ($\mu > 1$) phases:

\vspace{0.15cm}
\noindent
(i) When $\mu\leq1$, there is only the removable singularity at $k\rightarrow0^+$ as given by Eq.~(\ref{eq:smallkexpansion1}).
As a consequence, with the result $\int_{0}^{\infty}\,dx\,\sin(x)/x=\pi/2$, $|g(n)|$ is seen to approach the value $(\mu+1)/2$ for large $n$, 
as shown in panels \textbf{(b)} and \textbf{(c)} of Fig.~\ref{Figure-5}.

\vspace{0.15cm}
\noindent
(ii) When $\mu>1$, there is an additional removable singularity at $k\rightarrow~\pi^-$ corresponding to the first line of Eq.~\eqref{eq:smallkexpansion2}. 
By noting that $2\pi|g(n)|$ can be written equivalently as 
$\int_{0}^{\pi}dk\,\sin[n(\pi-k)] \, \mathcal{G}(\pi-k) = {(-1)}^{n+1}\int_{0}^{\pi}dk\,\sin(nk) \, \mathcal{G}(\pi-k) = {(-1)}^{n+1}n^{-1}\int_{0}^{n\pi}dp\,\sin(p) \, \mathcal{G}(\pi-p/n)$, for $k\rightarrow\pi^-$ the integral equals ${(-1)}^{n+1}\pi(\mu-1)$. 
The sum of the two contributions to the integral (from the $k\rightarrow0^+$ and $k\rightarrow\pi^-$ singularities) then yields $\pi(\mu+1)+{(-1)}^{n+1}\pi(\mu-1)$, which equals $2\pi$ for even $n$ and $2\pi\mu$ for odd $n$.
This double value $|g(n)|$ for large $n$ is evident in panel \textbf{(a)} of Fig.~\ref{Figure-5}.

\vspace{0.1cm}
All in all, the asymptotic behavior of $|g(n)|$ for large $n$ can be written in the compact form:
\begin{equation}
|g(n)|\simeq\left\{\begin{array}{ll}
\frac{1}{2}[(\mu+1)+{(-1)}^{n+1}(\mu-1)] & \quad (\mu>1) \vspace{0.2cm}\\
\frac{1}{2}(\mu+1) & \quad (\mu\leq1)
\end{array}\right. \,\, .
\label{general-asymptotic-result-g}
\end{equation} 
We have verified that a similar behavior is obtained more generally also when $2 \Delta \ne 1$.

\subsection{Extracting the exponential or power-law behavior of \\ the pair wave function}
\label{sec:numerics4}
It was concluded in Ref.~\cite{Neupert-2020} that, for a one-dimensional p-wave superconductor, the wave function of a Cooper pair falloffs exponentially with separation in the trivial phase and polynomially in the topological phase, 
in close analogy with the corresponding behavior of the Wannier-like functions.
In Ref.~\cite{Neupert-2020} this result was demonstrated ``by contradiction'' by relying on the properties of the Berry phase mentioned before, without, however, reporting explicit numerical calculations
for the pair wave function on the basis of its definition (\ref{definition-g}).
The results for $|g(n)|$ that we have reported both by extensive numerical calculations of Eq.~(\ref{definition-g}) [also in conjunction with the asymptotic estimates~(\ref{general-asymptotic-result-g})
and the analytic outcomes of the examples], appear to contradict both a mere exponential falloff of $|g(n)|$ in the trivial phase for $\mu > 1$ and a power-law falloff in the non-trivial phase for $\mu \le 1$ 
that were obtained in Ref.~\cite{Neupert-2020}.
It is thus worth delving into this issue so as to settle the above contradiction.

\begin{figure*}[t]
\centering
\includegraphics[width=18cm,angle=0]{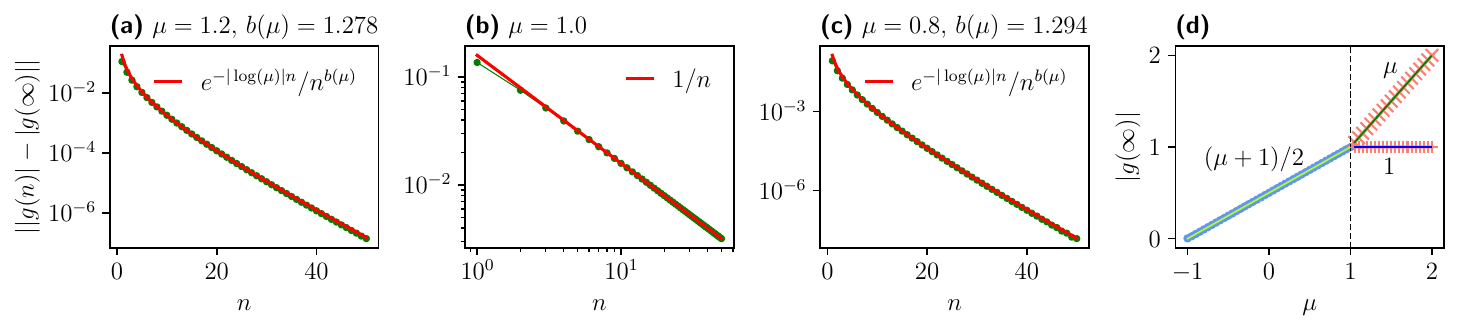}
\caption{Dependence on the lattice site $n$ of the pair wave function $|g(n)|$ of Eq.~\eqref{definition-g}, once the asymptotic behavior~(\ref{general-asymptotic-result-g}) has been suitably subtracted off as explained in the text.
              The cases with $\mu >1$, $\mu = 1$, and $\mu <1$ are reported separately in the three panels from \textbf{(a)} to \textbf{(c)}, respectively. Panel \textbf{(d)} shows the large-$n$ value of $|g(n)|$ for several values of $\mu$. 
              Numerical points (blue dots for $\mu<1$, and red pluses and crossed for $\mu>1$) are well overlapped to the expected asymptotic value from Eq.~\eqref{general-asymptotic-result-g}, which is $|g(\infty)|=(\mu+1)/2$ for $\mu<1$ (light-green line), 
              while for $\mu>1$ one has the alternating values $|g(\infty)|=1$ for even $n$ (dark-blue line) and $|g(\infty)|=\mu$ for odd $n$ (dark-green line). 
              The QCP value $\mu=1$ separating the two behaviors is marked by a vertical dashed black line.}
\label{Figure-6}
\end{figure*} 

To highlight the rate of falloff of $|g(n)|$ to the constant values (\ref{general-asymptotic-result-g}) for large $n$ in the various cases, we report in panels from \textbf{(a)} to \textbf{(c)} of Fig.~\ref{Figure-6} the same numerical data shown in panels from \textbf{(a)} to \textbf{(c)} of Fig.~\ref{Figure-5}, 
from which the appropriate asymptotic values~(\ref{general-asymptotic-result-g}) are now subtracted off, in such a way that the subtracted function $\tilde{g}(n)\coloneqq||g(n)|-|g(\infty)||$, where $|g(\infty)|\coloneqq\lim_{n\rightarrow\infty}|g(n)|$, converges eventually to zero for large $n$.
By numerical fits of the subtracted function $\tilde{g}(n)$ for large $n$, for \emph{either\/} $\mu > 1$ \emph{or\/} $\mu <1$ we find an exponential convergence of the type $e^{-|\log(\mu)|n}/n^{b(\mu)}$ 
for both even and odd $n$, although with the exponent $b(\mu)$ in the denominator that depends on $\mu$, while for $\mu = 1$ we find a simple $1/n$ power-law dependence.
This kind of behavior (from exponential, to power law, and then back to exponential across a critical point) is reminiscent of that occurring for standard correlation functions in a ``classical'' phase transition by the action of temperature \cite{LeBellac-1995}.

What is novel here is that in the present case, when crossing the QCP associated to a one-dimensional p-wave fermionic superfluid at zero temperature, is the fact that for large $n$ the pair wave function $|g(n)|$ converges to a single asymptotic value 
in the non-trivial (topological) phase when $\mu < 1$, while in the trivial phase when $\mu > 1$ two asymptotic values are alternatively reached for even and odd $n$.
This peculiar result is summarized in panel \textbf{(d)} of Fig.~\ref{Figure-6}, where the expected asymptotic behavior (\ref{general-asymptotic-result-g}) is recovered by our numerical calculations for all values of $\mu$.
Accordingly, we may conclude that the emergence of the non-trivial (topological) phase of a one-dimensional p-wave fermionic superfluid across the QCP at $\mu = 1$ is characterized by the lumping into a single value of the two asymptotic values of the pair wave function which occur in the trivial phase.
Even in this case have verified that a similar behavior is obtained quite generally also when $2 \Delta \ne 1$.

The case with $\mu > 1$ is worthy of further consideration, about the reaching of the non-interacting limit when $\mu$ increases while keeping $\Delta$ fixed, as we do in the present context.
By the action of the ``ghost'' band mentioned before, $\mu$ can be made to increase in such a way that $\epsilon(k) \rightarrow |\xi(k)|$ from Eq.~(\ref{epsilon}) with $\xi(k) < 0$,
and that $|u(k)| \rightarrow 0$ from Eq.~(\ref{u-notation}) and $|v(k)| \rightarrow 1$ from Eq.~(\ref{v-notation}).
Within the present constraints, the non-interacting limit is effectively reached by an imbalance of the probability amplitudes for occupying nearest-neighbor sites, to which the quantity $\Delta$ of Eq.~(\ref{local-gap-definition})
has been limited from the outset.
At the same time, for increasing $\mu$ the absolute values of the correlation function $f(n)$ consistently decrease, as shown in panels from \textbf{(f)} to \textbf{(d)} of Fig.~\ref{Figure-5}.
This should explain on physical grounds the occurrence of the result~(\ref{general-asymptotic-result-g}), which distinguishes between nearest-neighbors (even and odd) sites. 

\section{Two coupled linear chains in the normal phase}
\label{sec:two-linear-chains}
The emergence of the QCP for the one-dimensional model system introduced before is due to the presence of the \emph{interaction} Hamiltonian, which gives rise to the mean-field term in 
Eq.~\eqref{system-Hamiltonian-interacting-2} and results in the non-trivial falloff of the Wannier-like functions. 
Here, we show that a similar phenomenology arises in a system of \emph{non-interacting} particles embedded in a $(1+\epsilon)$-dimensional space. 
In practice, this system can be realized by a \emph{two-leg ladder}, which consists of two coupled linear chains with both intra- and inter-leg hoppings. 
In this way, the particles states are identified by two quantum numbers, namely, the site position $n$ along the chain as before, and an additional quantum number that distinguishes between the two chains
(and plays the role of an additional ``synthetic'' dimension - cf. Refs.~\cite {Fallani-2015} and \cite{MCS-2017}, and references therein).
No specific reference to the spin of the particles is required in the present case.

The present discussion parallels the content of the previous analysis on the one-dimensional p-wave fermionic superfluid, and considers the topological properties that emerge from the two-leg ladder model, further distinguishing the cases of s-vs-s and s-vs-p orbitals.
The latter case will result in the presence of \emph{two\/} QCPs, which a richer interplay between topological and non-topological phases (occurring now in the normal instead of the superfluid phase) with respect to the p-wave superfluid with only one QCP.

\subsection{Two-leg ladder model}
\label{sec:two-leg-ladder}
\begin{figure}[t]
\begin{center}
\includegraphics[width=8.4cm,angle=0]{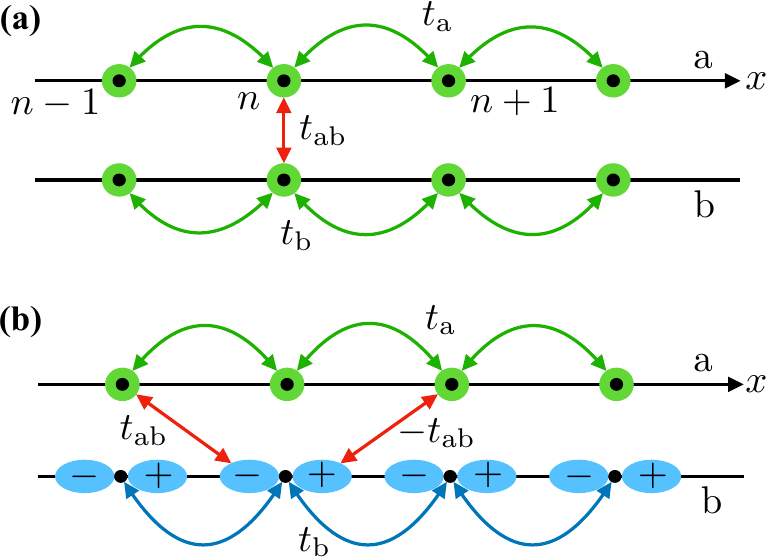}
\caption{\textbf{(a)} Two coupled linear chains (``a'' and ``b'') both with s-like orbitals (green circles), with intra-chain hopping integrals $t_{\rm a}>0$ and $t_{\rm b}>0$, and inter-chain hopping integral $t_{\rm ab}>0$ between sites with the same $n$.   
              \textbf{(b)} Two coupled linear chains with s-like orbitals (green circles) on chain ``a'' and p-like orbitals (blue ellipses) on chain ``b'', with intra-chain hopping integrals $t_{\rm a}>0$ and $t_{\rm b}<0$ 
              and inter-chain hopping integral $t_{\rm ab}$ between nearest-neighbor sites $n-1$ and $n$, and $-t_{\rm ab}$ between sites $n$ and $n+1$. 
              Here, the different sign in front of the hopping integral $t_{\rm ab}$ stems from the different sign of the overlap between s-like and p-like orbitals in the two cases.}
\label{Figure-7}
\end{center} 
\end{figure}

The two-leg ladder model consists of non-interacting spin-less fermions embedded in two linear chains, identified as ``a'' and ``b''. 
As before, to the lattice sites in each chain there corresponds an integer $n$, with the lattice constant set equal to unity.
Hopping processes result from both intra-chain hopping along each chain with integrals $t_{\rm a}$ and $t_{\rm b}$, and inter-chain hopping between the two chains with integral $t_{\rm ab}$ (see Fig.~\ref{Figure-7}). 

In this case, only the one-body term~\eqref{system-Hamiltonian-noninteracting-1} of the Hamiltonian is active, with the field operator $\Psi(x)$ represented in the form
\begin{equation}
\Psi(x) = \sum_{n} \, \left( \phi_{\mathrm{a}}(x-n) \, c_{\mathrm{a}}(n) + \phi_{\mathrm{b}}(x-n) \, c_{\mathrm{b}}(n) \right)
\label{field-operator-two-coupled-chains-1}
\end{equation}
in the place of Eq.~\eqref{one-band-approximation}, where $\phi_{\mathrm{a}}(x)$ and $\phi_{\mathrm{b}}(x)$ are real orbitals.
The one-body Hamiltonian then reads
\begin{eqnarray}
H_{1} & = & \sum_{n_{1} n_{2}} \left( \! c^{\dagger}_{\mathrm{a}}(n_{1}) c_{\mathrm{a}}(n_{2}) \!\! \int \!\! dx \, \phi_{\mathrm{a}}^{*}(x-n_{1}) \, h \, \phi_{\mathrm{a}}(x-n_{2}) \right.
\nonumber \\
&& +  c^{\dagger}_{\mathrm{b}}(n_{1}) c_{\mathrm{b}}(n_{2}) \!\! \int \!\! dx \, \phi_{\mathrm{b}}^{*}(x-n_{1}) \, h \, \phi_{\mathrm{b}}(x-n_{2})
\nonumber \\
&& +  c^{\dagger}_{\mathrm{a}}(n_{1}) c_{\mathrm{b}}(n_{2}) \!\! \int \!\! dx \, \phi_{\mathrm{a}}^{*}(x-n_{1}) \, h \, \phi_{\mathrm{b}}(x-n_{2})
\nonumber \\
&& +  \left. c^{\dagger}_{\mathrm{b}}(n_{1}) c_{\mathrm{a}}(n_{2}) \!\! \int \!\! dx \, \phi_{\mathrm{b}}^{*}(x-n_{1}) \, h \, \phi_{\mathrm{a}}(x-n_{2}) \right) ,
\label{two-linear-chains-Hamiltonian-1}
\end{eqnarray} 
where we distinguish the two cases depicted in Fig.~\ref{Figure-7}, when both chains ``a'' and ``b'' host s-like orbitals and when chain ``a'' hosts s-like orbitals while chain ``b'' hosts p-like orbitals.

\begin{figure}[t]
\begin{center}
\includegraphics[width=8.0cm,angle=0]{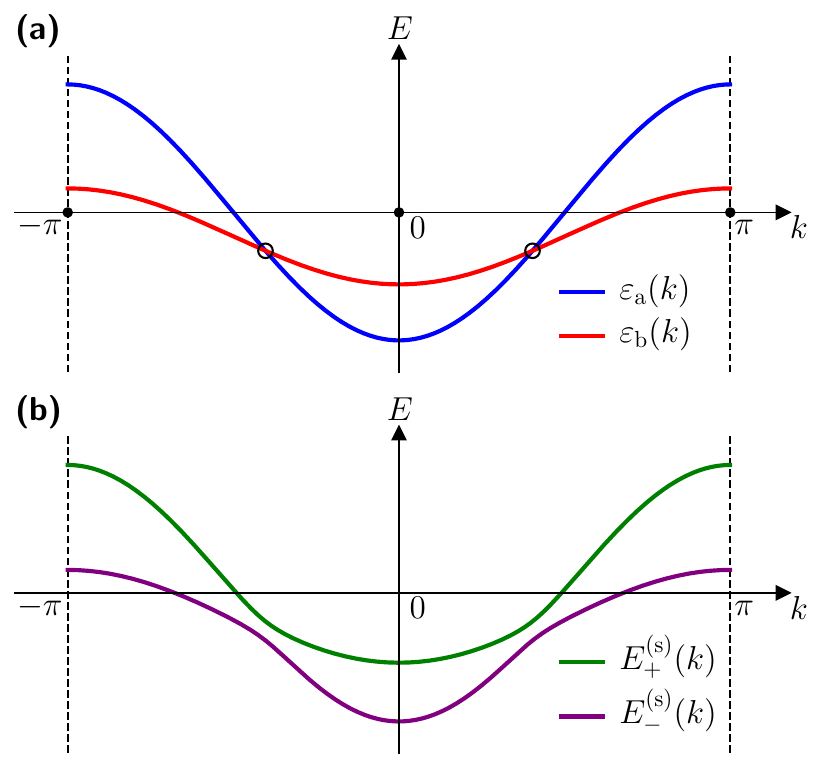}
\caption{\textbf{(a)} Dispersions $\varepsilon_{\mathrm{a}}(k)$ (with $t_{\rm a}>0$ and $\mu_{\rm a}=0$) and $\varepsilon_{\mathrm{b}}(k)$ (with $t_{\rm b}>0$ and $\mu_{\rm b}=t_{\rm b}$) of the two bands 
              given by Eq.~\eqref{two-linear-chains-Hamiltonian-201} before inter-chain hybridization.
              Here, the circles mark the values $\pm k_x$ where the band crossings occur, as obtained from Eq.~\eqref{eq:positionkcrossingpoint01}.
              \textbf{(b)} Dispersions of the two bands from Eq.~\eqref{eigenvalues_chains} resulting when the inter-chain hopping $t_{\rm ab}>0$ is active, such that the crossings of panel \textbf{(a)} turn into avoided crossings. 
              The numerical parameters are $t_{\rm a}=0.4$, $t_{\rm b}=0.15$, $\mu_{\rm a}=0$, $\mu_{\rm b}=t_{\rm b}$, and $t_{\rm ab}=0.06$.}
\label{Figure-8}
\end{center} 
\end{figure} 

\subsection{s-like vs s-like orbitals}
\label{sec:s-vs-s-leg-ladder}
When both chains host s-like orbitals (like in panel \textbf{(a)} of Fig.~\ref{Figure-7}), in the first two (intra-chain) terms of Eq.~\eqref{two-linear-chains-Hamiltonian-1} we take $n_{1} = (n_{2},n_{2} \pm1)$ while in the last two (inter-chain) terms we take $n_{1} = n_{2}$. 
We then define
\begin{equation}
\mu_{\mathrm{a}/\mathrm{b}} = -\! \int \!\! dx \, \phi_{\mathrm{a}/\mathrm{b}}^{*}(x) \, h \, \phi_{\mathrm{a}/\mathrm{b}} (x)
\label{on-site-energy}
\end{equation}
for the on-site energy (which may be different in the two cases), 
\begin{equation}
t_{\mathrm{a}/\mathrm{b}} = - \! \int \!\! dx \, \phi_{\mathrm{a}/\mathrm{b}}^{*}(x-1) \, h \, \phi_{\mathrm{a}/\mathrm{b}} (x) 
\label{intra-chain-hoppings}
\end{equation}
for the intra-chain hopping integrals within each chain, and
\begin{equation}
t_{\mathrm{a}\mathrm{b}} = - \! \int \!\! dx \, \phi_{\mathrm{a}/\mathrm{b}}^{*}(x) \, h \, \phi_{\mathrm{b}/\mathrm{a}} (x)
\label{inter-chain-hopping}
\end{equation}
for the inter-chain hopping integral between chains. 
With the lattice Fourier transform~\eqref{lattice-Fourier-transform}, the Hamiltonian in Eq.~\eqref{two-linear-chains-Hamiltonian-1} then becomes
\begin{equation}
H_{1} = \sum_{k}^{BZ} \left(\begin{array}{cc} c_{\mathrm{a}}(k)^{\dagger}&c_{\mathrm{b}}(k)^{\dagger} \end{array}\right) \!
\left( \hspace{-0.1cm} 
\begin{array}{cc}
\varepsilon_{\mathrm{a}}(k)    &  - t_{\mathrm{a}\mathrm{b}}  \\
- t_{\mathrm{a}\mathrm{b}}     & \varepsilon_{\mathrm{b}}(k)
\end{array}
\hspace{-0.1cm} \right) \! 
\left( \hspace{-0.1cm} 
\begin{array}{c}
c_{\mathrm{a}}(k)   \\
c_{\mathrm{b}}(k)     
\end{array}
\hspace{-0.1cm} \right) \,\, ,
\label{two-linear-chains-Hamiltonian-2}
\end{equation}
where
\begin{equation}
\varepsilon_{\mathrm{a}/\mathrm{b}}(k) = - 2 \, t_{\mathrm{a}/\mathrm{b}} \cos (k) - \mu_{\mathrm{a}/\mathrm{b}} 
\label{two-linear-chains-Hamiltonian-201}
\end{equation}
are the band dispersions of each chain before hybridization by the inter-chain hopping $t_{\rm ab}$
(from which the thermodynamic chemical potential $\mu$ is meant to be subtracted off once the hybridization by $t_{\rm ab}>0$ is active).
We are thus left with diagonalizing the matrix
\begin{equation}
M^{(\mathrm{s-s})}(k) = 
\left( \hspace{-0.1cm} 
\begin{array}{cc}
\varepsilon_{\mathrm{a}}(k)    &  - t_{\mathrm{a}\mathrm{b}}  \\
- t_{\mathrm{a}\mathrm{b}}     & \varepsilon_{\mathrm{b}}(k)
\end{array}
\hspace{-0.1cm} \right) \, ,
\label{M_chains-ss-matrix}
\end{equation}
yielding
\begin{equation}
\left( \hspace{-0.1cm} 
\begin{array}{cc}
\varepsilon_{\mathrm{a}}(k)    &  - t_{\mathrm{a}\mathrm{b}}  \\
- t_{\mathrm{a}\mathrm{b}}     & \varepsilon_{\mathrm{b}}(k)
\end{array}
\hspace{-0.1cm} \right) \! 
\left( \hspace{-0.1cm} 
\begin{array}{c}
\psi_{\rm a}(k)  \\
\psi_{\rm b}(k)     
\end{array}
\hspace{-0.1cm} \right)
= E^{\rm (s)}(k) \! 
\left( \hspace{-0.1cm} 
\begin{array}{c}
\psi_{\rm a}(k)  \\
\psi_{\rm b}(k)     
\end{array}
\hspace{-0.1cm} \right) 
\label{diagonalization-M_chains-matrix}
\end{equation}
with eigenvalues 
\begin{equation}
E^{\rm (s)}_\pm(k) = \left( \frac{\varepsilon_{\mathrm{a}}(k) + \varepsilon_{\mathrm{b}}(k)}{2} \right) \pm \sqrt{ \left( \frac{\varepsilon_{\mathrm{a}}(k) - \varepsilon_{\mathrm{b}}(k)}{2} \right)^{2} + t_{\mathrm{a}\mathrm{b}}^{2}} 
\label{eigenvalues_chains}
\end{equation}
and the normalization condition
\begin{equation}
|\psi_{\rm a}(k)|^{2} + |\psi_{\rm b}(k)|^{2} = 1 
\label{normalization-condition_chains}
\end{equation}
for each $k$ in the BZ. 

A typical ``crossing'' configuration of the two bands $\varepsilon_{\mathrm{a}}(k)$ and $\varepsilon_{\mathrm{b}}(k)$ of Eq.~\eqref{two-linear-chains-Hamiltonian-201} 
(before hybridization by the inter-chain hopping $t_{\mathrm{a}\mathrm{b}}$) is shown in panel \textbf{(a)} of Fig.~\ref{Figure-8}. 
Here, the values $\pm k_{\rm x}$ where the band crossing $\varepsilon_{\mathrm{a}}(k_{\rm x})=\varepsilon_{\mathrm{b}}(k_{\rm x})$ occurs are given by
\begin{equation}
\cos(k_{\rm x})=-\frac{1}{2}\,\frac{\mu_{\rm a}-\mu_{\rm b}}{t_{\rm a}-t_{\rm b}} \,\, ,
\label{eq:positionkcrossingpoint01}
\end{equation}
which has solution as long as $|\cos(k_{\rm x})|<1$, that is when $|\mu_{\rm a}-\mu_{\rm b}|<2|t_{\rm a}-t_{\rm b}|$. 
On physical grounds, this crossing is possible because the two uncoupled chains constitute a \emph{quasi-one-dimensional} system, to the extent that the on-site energies $\varepsilon_{\mathrm{a}/\mathrm{b}}$ 
as well as the intra-chain hoppings $t_{\mathrm{a}/\mathrm{b}}$ are meant to be varied independently.

This crossing does not survive after the hybridization due to the inter-chain hopping $t_{\rm ab}$, as shown in panel \textbf{(b)} of Fig.~\ref{Figure-8}.
However, in this case no singularity occurs in the eigenvectors of Eq.~(\ref{diagonalization-M_chains-matrix}), similarly to the case of truly one-dimensional bands discussed in detail in Ref.~\cite{Kohn-1959}.
In the present case, this feature can be ascribed to the fact that the off-diagonal term of the matrix (\ref{M_chains-ss-matrix}) does not depend on the wave vector $k$, a limitation which we now remove in the following case.

\subsection{s-like vs p-like orbitals}
\label{sec:s-vs-p-leg-ladder}
When chain ``a'' hosts s-like orbitals and chain ``b'' hosts p-like orbitals, like in panel \textbf{(b)} of Fig.~\ref{Figure-7}, the system Hamiltonian is still of the form~\eqref{two-linear-chains-Hamiltonian-2}, but now with the s-p matrix
\begin{equation}
M^{(\mathrm{s-p})}(k) = 
\left( \hspace{-0.1cm} 
\begin{array}{cc}
\varepsilon_{\mathrm{a}}(k)                   &  - 2 i t_{\mathrm{a}\mathrm{b}} \sin (k) \\
2 i t_{\mathrm{a}\mathrm{b}} \sin (k)      &  \varepsilon_{\mathrm{b}}(k)
\end{array}
\hspace{-0.1cm} \right) 
\label{M_chains-sp-matrix}
\end{equation}
in the place of the s-s matrix~\eqref{M_chains-ss-matrix}.
In this case, 
\begin{equation}
- t_{\mathrm{a}} = \! \int \!\!\! dx \, \phi_{\mathrm{a}}(x-1) \, h \, \phi_{\mathrm{a}}(x) =  \! \int \!\!\! dx \, \phi_{\mathrm{a}}(x+1) \, h \, \phi_{\mathrm{a}}(x) < 0 
\label{intra-chain-hopping-a}
\end{equation}
is the s-s hopping for chain ``a'',
\begin{equation}
- t_{\mathrm{b}} = \! \int \!\!\! dx \, \phi_{\mathrm{b}}(x-1) \, h \, \phi_{\mathrm{b}}(x) =  \! \int \!\!\! dx \, \phi_{\mathrm{b}}(x+1) \, h \, \phi_{\mathrm{b}}(x) > 0
\label{intra-chain-hopping-b}
\end{equation}
the p-p hopping for chain ``b'', and
\begin{equation}
t_{\mathrm{ab}} = \! \int \!\!\! dx \, \phi_{\mathrm{a}}(x-1) \, h \, \phi_{\mathrm{b}}(x) = - \, \! \int \!\!\! dx \, \phi_{\mathrm{a}}(x+1) \, h \, \phi_{\mathrm{b}}(x) > 0
\label{inter-chain-hopping-ab}
\end{equation}
the s-p hopping between nearest-neighbor sites on the two chains.
(The signs of the terms (\ref{intra-chain-hopping-a}), (\ref{intra-chain-hopping-b}), and (\ref{inter-chain-hopping-ab}) are meant to correspond to those of the hopping integrals that describe, e.g., the s-p bands of NaCl near the energy gap.)
On the other hand, the inter-chain on-site term 
\begin{equation}
\int \!\!\! dx \, \phi_{\mathrm{a}}(x) \, h \, \phi_{\mathrm{b}}(x) =  0 
\label{on-site-a-b}
\end{equation}
vanishes by symmetry. 

The matrix $M^{(\mathrm{s-p})}(k)$ of Eq.~\eqref{M_chains-sp-matrix} has a form similar to that of the matrix $M(k)$ of Eq.~\eqref{M-matrix}, with the difference that the diagonal elements now correspond to the bands of two distinct chains with opposite curvature 
and are thus not simply related to each other like the diagonal elements in Eq.~\eqref{M-matrix},
while the off-diagonal elements are obtained by replacing $\Delta(k) = - 2 i \Delta \sin(k)$ of Eq.~\eqref{k_dependent-gap-definition} by $\tau(k)=-2\,i\,t_{\rm ab}\sin(k)$.

Accordingly, the eigenvalues of the matrix~\eqref{M_chains-sp-matrix} 
\begin{equation}
E^{\rm(p)}_\pm(k) = \frac{\varepsilon_{\rm a}(k)+\varepsilon_{\rm b}(k)}{2}\pm\sqrt{{\left(\frac{\varepsilon_{\rm a}(k)-\varepsilon_{\rm b}(k)}{2}\right)}^2+{|\tau(k)|}^2} 
\label{eq:spchain1}
\end{equation}
differ from the eigenvalues $E^{\rm (s)}_\pm(k)$ of Eq.~\eqref{eigenvalues_chains} by the replacement $t_{\mathrm{ab}}^{2} \, \rightarrow \, |\tau(k)|^2=4 t_{\mathrm{ab}}^{2} \sin^{2}(k)$ (besides the difference in sign of $t_{\rm b}$).
These differences considerably modify the topological properties of the two (s-s and s-p) cases.

\subsection{Mapping between the two-leg s-p ladder \\ and the p-wave superfluid chain}
\label{sec:mapping-two-legs-vs-superfluid}
The above similarity, between the formal structure of the matrices~\eqref{M-matrix} and~\eqref{M_chains-sp-matrix}, can further be exploited so as to establish a \emph{formal mapping\/} between the corresponding eigenvectors.
This task can be accomplished as follows.

The expression of the eigenvectors of the matrix~\eqref{M_chains-sp-matrix}, namely, 
\begin{equation}
\left(\!\! \begin{array}{c}\psi^{(\pm)}_{\rm a}(k)\\\psi^{(\pm)}_{\rm b}(k)\end{array} \!\! \right) = \left( \!\! \begin{array}{c}1\\ \frac{\tau(k)^*}{E^{\rm(p)}_\pm(k)-\varepsilon_{\rm b}(k)} \end{array} \!\! \right) \,
                                                                                                                                                  \frac{1}{\sqrt{1 + \frac{|\tau(k)|^2}{(E^{\rm(p)}_\pm(k)-\varepsilon_{\rm b}(k))^2}}}
\label{eq:spchain2}
\end{equation}
can be manipulated by introducing the quantities
\begin{eqnarray}
\xi_{\rm c}(k) = \frac{\varepsilon_{\rm a}(k)-\varepsilon_{\rm b}(k)}{2}&=&-\left(t_{\rm a}-t_{\rm b}\right)\cos(k)-\frac{\mu_{\rm a}-\mu_{\rm b}}{2} 
\nonumber\\
&=&-2t_{\rm c}\cos(k)-\mu_{\rm c}
\label{eq:spchain2x1}
\end{eqnarray}
where
\begin{equation}
t_{\rm c}=\frac{t_{\rm a}-t_{\rm b}}{2} \hspace{0.8cm} , \qquad \mu_{\rm c}=\frac{\mu_{\rm a}-\mu_{\rm b}}{2}
\label{eq:spchain2x1bis0}
\end{equation}
(with $t_{\rm c}>0$ since $t_{\rm a}>0$ and $t_{\rm b}<0$), and
\begin{equation}
\epsilon_{\rm c}(k) = \sqrt{\xi_{\rm c}^2(k)+{|\tau(k)|}^2} \,\, .
\label{eq:spchain3bis01}
\end{equation} 
In this way, from Eq.~(\ref{eq:spchain1}) we obtain $E^{\rm(p)}_\pm(k) - \varepsilon_{\rm b}(k) = \xi_{\rm c}(k) \pm \epsilon_{\rm c}(k)$, such that the expression within square root in Eq.~(\ref{eq:spchain2}) becomes
\begin{equation}
1 + \frac{ |\tau(k)|^2 }{ ( E^{\rm(p)}_\pm(k) - \varepsilon_{\rm b}(k) )^2 } = \frac{2 \, \epsilon_{\rm c}(k)}{\epsilon_{\rm c}(k) \pm \xi_{\rm c}(k)} \, .
\label{manipulation-square-root}
\end{equation}
By a similar token, we obtain for the factor in the second component of the column vector on the right-hand side of Eq.~(\ref{eq:spchain2})
\begin{equation}
\frac{\tau(k)^*}{E^{\rm(p)}_\pm(k)-\varepsilon_{\rm b}(k)} = \frac{\tau(k)^*}{\xi_{\rm c}(k) \pm \epsilon_{\rm c}(k)} = \frac{\tau(k)^*}{|\tau(k)|} \, \frac{ \sqrt{\epsilon_{\rm c}(k)^2 - \xi_{\rm c}(k)^2} }{\xi_{\rm c}(k) \pm \epsilon_{\rm c}(k)}
\label{manipulation-factor-second-component}
\end{equation}
where $\frac{\tau(k)^*}{|\tau(k)|} = i \, \mathrm{sign}(k)$ and $|\tau(k)| = \sqrt{\epsilon_{\rm c}(k)^2 - \xi_{\rm c}(k)^2}$ from Eq.~(\ref{eq:spchain3bis01}).
Entering the results (\ref{manipulation-square-root}) and (\ref{manipulation-factor-second-component}) into the expression (\ref{eq:spchain2}) of the ($\pm$) eigenvectors, we obtain eventually
(with an independent rearrangement of the overall factors for the two eigenvectors)
\begin{subequations}
\begin{align}
\left(\begin{array}{c}\psi^{(+)}_{\rm a}(k)\\\psi^{(+)}_{\rm b}(k)\end{array}\right)&=\left(\begin{array}{c} {\rm sign}(k) |u_{\rm c}(k)|\\i\,|v_{\rm c}(k)|\end{array}\right) \\
\left(\begin{array}{c}\psi^{(-)}_{\rm a}(k)\\\psi^{(-)}_{\rm b}(k)\end{array}\right)&=\left(\begin{array}{c}i\,|v_{\rm c}(k)|\\{\rm sign}(k)\,|u_{\rm c}(k)|\end{array}\right) 
\end{align}
\label{eq:spchain12}
\end{subequations}

\noindent
in analogy with the expressions (\ref{+eigenvector-final}) and (\ref{-eigenvector-final}), where we have introduced the notation
\begin{subequations}
\begin{align}
|u_{\rm c}(k)|&=\sqrt{\frac{1}{2}\left(1+\frac{\xi_{\rm c}(k)}{\epsilon_{\rm c}(k)}\right)}\\
|v_{\rm c}(k)|&=\sqrt{\frac{1}{2}\left(1-\frac{\xi_{\rm c}(k)}{\epsilon_{\rm c}(k)}\right)} 
\end{align}
\label{eq:spchain11}
\end{subequations}
in analogy with the expressions (\ref{u-notation}) and (\ref{v-notation}).

With the definitions~\eqref{eq:spchain2x1bis0}, before hybridization the two bands $\varepsilon_{\mathrm{a}}(k) $ and $\varepsilon_{\mathrm{b}}(k)$ would cross for $\cos(k_{\rm x}) = -\mu_{\rm c}/2t_{\rm c}$ 
as long as $|\mu_{\rm c}| < 2 |t_{\rm c}| = 2t_{\rm c}$ since $t_{\rm c}>0$.
This identifies the two (lower and upper) boundaries
\begin{equation}
\mu_1=-2t_{\rm c} \qquad \mu_2=2t_{\rm c} 
\label{eq:spchain12bis01}
\end{equation}
with $\mu_1<\mu_2$, each of which is associated with a QCP where a phenomenology similar to that discussed for the one-dimensional p-wave fermionic superfluid occurs, as shown below. 

\subsection{Wannier functions}
\label{sec:Wannier-functions}
 There remains to consider the Wannier functions for the s-p ladder which are separately associated with the bands of higher and lower energies at each $k$, given respectively by $E^{\rm(p)}_{+}(k)$ and $E^{\rm(p)}_{-}(k)$ of Eq.~(\ref{eq:spchain1}).
In this case, the topological obstruction for the Wannier functions is expected to manifests itself when crossing \emph{each\/} of the two QCPs associated with these bands.
It is thus worth showing how these functions can be related to Wannier-like lattice functions like those introduced and numerically evaluated for the p-wave fermionic superfluid in Eqs.~(\ref{Wannier-like-function-1}) and (\ref{Wannier-like-function-2}).

With the system Hamiltonian (\ref{two-linear-chains-Hamiltonian-1}) expressed in second quantization, it is convenient to manipulate the field operator (\ref{field-operator-two-coupled-chains-1}) for the coupled chains in the following way
\begin{equation}
\Psi(x) = \sum_{k}^{BZ} \left( \Phi_{k,{\rm a}}(x) \, c_{k,{\rm a}} + \Phi_{k,{\rm b}}(x) \, c_{k,{\rm b}} \right) \, ,
\label{field-operator-two-coupled-chains-2}
\end{equation}
where we have utilized the lattice Fourier transform (\ref{lattice-Fourier-transform}) for each orbital and introduced the Bloch sums
\begin{equation}
\Phi_{k,{\rm a/b}}(x) = \frac{1}{\sqrt{N}} \sum_{n} e^{i k n} \, \phi_{\mathrm{a}/\mathrm{b}}(x-n) \, .
\label{Bloch-sums}
\end{equation}
Next, we introduce the (unitary) matrix
\begin{equation}
U(k) = 
\left( \hspace{-0.1cm} 
\begin{array}{cc}
\psi_{\mathrm{a}}^{(+)}(k)   &  \psi_{\mathrm{a}}^{(-)}(k) \\
\psi_{\mathrm{b}}^{(+)}(k)   &  \psi_{\mathrm{b}}^{(-)}(k)
\end{array}
\hspace{-0.1cm} \right) 
\label{U-matrix}
\end{equation}
in terms of the band eigenvectors \eqref{eq:spchain12}, such that the field operator (\ref{field-operator-two-coupled-chains-2}) can be brought to the form:
\begin{eqnarray}
\Psi(x) & = & \sum_{k}^{BZ} \left( \Phi_{k,{\rm a}}(x),\Phi_{k,{\rm b}}(x) \right) U(k) \, U^{\dagger}(k) 
\left( \hspace{-0.1cm} 
\begin{array}{c}
c_{k,\mathrm{a}}   \\
c_{k,\mathrm{b}}     
\end{array}
\hspace{-0.1cm} \right)
\nonumber \\
& = & \sum_{k}^{BZ} \left( \Phi^{(+)}_{k}(x),\Phi^{(-)}_{k}(x) \right) \!
\left( \hspace{-0.1cm} 
\begin{array}{c}
\alpha^{(+)}_k   \\
\alpha^{(-)}_k 
\end{array}
\hspace{-0.1cm} \right)
\nonumber \\
& = & \sum_{n} \left( \!\! \frac{1}{\sqrt{N}} \sum_{k}^{BZ} e^{-i n k} \, \Phi^{(+)}_{k}(x) \!\! \right) \left( \!\! \frac{1}{\sqrt{N}} \sum_{k'}^{BZ} e^{i n k'} \, \alpha^{(+)}_{k'} \!\! \right)
\nonumber \\
& + & \sum_{n} \left( \!\! \frac{1}{\sqrt{N}} \sum_{k}^{BZ} e^{-i n k} \, \Phi^{(-)}_{k}(x) \!\! \right) \left( \!\! \frac{1}{\sqrt{N}} \sum_{k'}^{BZ} e^{i n k'} \, \alpha^{(-)}_{k'} \!\! \right)
\nonumber \\
& = & \sum_{n} \left( w^{(+)}(x-n) \, \alpha^{(+)}_{n} + w^{(-)}(x-n) \, \alpha^{(-)}_{n} \right) \, .
\label{field-operator-two-coupled-chains-3}
\end{eqnarray}
In the above expression, we have:
\vspace{0.1cm}

\vspace{0.1cm}
\noindent
(i) Introduced the real-space eigenfunctions $\Phi^{(\pm)}_{k}(x)$ for the two bands, given by
\begin{equation}
\left(\begin{array}{cc}\Phi^{(+)}_{k}(x)&\Phi^{(-)}_{k}(x) \end{array}\right) = \left(\begin{array}{cc} \Phi_{k,\mathrm{a}}(x)&\Phi_{k,\mathrm{b}}(x) \end{array}\right) U(k) \,\, ;
\label{real-space-band-eigenfunctions}
\end{equation}

\vspace{0.1cm}
\noindent
(ii) Defined the quasi-particle operators in analogy to Eq.~(\ref{Bogoliubov-Valatin-operators-1})
\begin{equation}
\left( \hspace{-0.1cm} 
\begin{array}{c}
\alpha^{(+)}_k  \\
\alpha^{(-)}_k      
\end{array}
\hspace{-0.1cm} \right)
 =  U^{\dagger}(k) 
\left( \hspace{-0.1cm} 
\begin{array}{c}
c_{k,\mathrm{a}}   \\
c_{k,\mathrm{b}}     
\end{array}
\hspace{-0.1cm} \right) \,\, ,
\label{quasi-particle-operators}
\end{equation}
with the associated lattice Fourier transform
\begin{equation}
\alpha^{(\pm)}_{n} = \frac{1}{\sqrt{N}} \sum_{k}^{BZ} e^{i n k} \alpha^{(\pm)}_k \,\, ;
\label{lattice-Fourier-transform-quasi-particle}
\end{equation}

\vspace{0.1cm}
\noindent
(iii) Introduced the (real space) Wannier functions 
\begin{equation}
w^{(\pm)}(x-n) = \frac{1}{\sqrt{N}} \sum_{k}^{BZ} e^{-i n k} \Phi^{(\pm)}_{k}(x) 
\label{corresponding-Wannier-functions}
\end{equation}
associated with the band eigenfunctions (\ref{real-space-band-eigenfunctions}), which amounts to the canonical definition of the Wannier functions. 

For computational purposes, it is preferable to cast the expression (\ref{corresponding-Wannier-functions}) in an alternative form, by expressing the band eigenfunctions therein
in terms of the Bloch sums like in Eq.~(\ref{real-space-band-eigenfunctions}), thus writing
\begin{eqnarray}
w^{(\pm)}(x-n)&=&\sum_{n'}\left(\varphi^{(\pm)}_{\mathrm{a}}(n-n')\,\phi_{\mathrm{a}}(x-n')\right.\nonumber\\
&&\hspace{0.3cm}+\left.\varphi^{(\pm)}_{\mathrm{b}}(n-n')\,\phi_{\mathrm{b}}(x-n')\right)
\label{eq:numericaltestwannierfunction20}
\end{eqnarray}
in terms of the Wannier lattice functions
\begin{subequations}
\begin{align}
\varphi^{(\pm)}_{\mathrm{a}}(n)&=\frac{1}{N}\sum_{k}^{BZ}e^{-ikn}\,\psi^{(\pm)}_{\mathrm{a}}(k) \\
\varphi^{(\pm)}_{\mathrm{b}}(n)&=\frac{1}{N}\sum_{k}^{BZ}e^{-ikn}\,\psi^{(\pm)}_{\mathrm{b}}(k) \,\, .
\end{align}
\label{eq:numericaltestwannierfunction21}
\end{subequations}
To the extent that the orbitals $\phi_{\mathrm{a/b}}(x-n')$ are spatially localized at about $x \simeq n'$, the ``localization'' at about $x \simeq n$ of the Wannier functions $w^{(\pm)}(x-n)$ given by Eq.~(\ref{eq:numericaltestwannierfunction20}) 
stems from the ``localization'' of the Wannier lattice functions (\ref{eq:numericaltestwannierfunction21}) at about $n \simeq n'$.

As a consequence, with the form \eqref{eq:spchain12} of the band eigenvectors $\psi^{(\pm)}_{\mathrm{a/b}}(k)$, we are again led to consider the Wannier-like lattice functions like in Eqs.~(\ref{Wannier-like-function-1}) and (\ref{Wannier-like-function-2}), thus defining
\begin{eqnarray}
\hspace{-0.5cm} w_{n}^{(c1)} \! & = & \! \frac{1}{N} \! \sum_{k}^{BZ} e^{i n k} {\rm sign} (k) |u_{c}(k)| = i \!\! \int_{0}^{\pi} \!\! \frac{d k}{\pi} \sin (kn)\,|u_{c}(k)| 
\label{Wannier-like-function-c1}  \nonumber\\\\
\hspace{-0.5cm}  w_{n}^{(c2)} \! & = & \! \frac{1}{N} \! \sum_{k}^{BZ} e^{i n k} i |v_{c}(k)| = i \!\! \int_{0}^{\pi}\frac{d k}{\pi} \cos (kn)\,|v_{c}(k)| 
\label{Wannier-like-function-c2} 
\end{eqnarray}
in terms of the factors (\ref{eq:spchain11}).
In addition, owing to the occurrence in the present case of two QCPs (instead of only one like for the p-wave fermionic superfluid), it turns out that in some regime of the parameter $\mu_{c}$ (when the system is in the trivial phase) discontinuities in the band eigenvectors \eqref{eq:spchain12} can effectively be eliminated by multiplying both components of these eigenvectors by ${\rm sign}(k)$, thereby interchanging the overall factor ${\rm sign}(k)$ between the two components of the expressions \eqref{eq:spchain12}.
Under these circumstances, we are led to consider two additional Wannier-like lattice functions, of the form
\begin{eqnarray}
w^{(c3)}_n \! & = & \! \int_{0}^{\pi}\frac{dk}{\pi}\,\cos(kn)\,|u_{\rm c}(k)| 
\label{Wannier-like-function-c3}  \\
w^{(c4)}_n \! & = & \! - \! \int_{0}^{\pi}\frac{dk}{\pi}\,\sin(kn)\,|v_{\rm c}(k)| \, ,
\label{Wannier-like-function-c4} 
\end{eqnarray}
to be utilized in the place of the expressions (\ref{Wannier-like-function-c1}) and (\ref{Wannier-like-function-c2}) to determine the spatial falloff of the Wannier function (\ref{eq:numericaltestwannierfunction20}).
As before, the factor $i$ in Eqs.~(\ref{Wannier-like-function-c1}) and (\ref{Wannier-like-function-c2}), as well as the minus sign in Eq.~(\ref{Wannier-like-function-c4}), will be eliminated from further
consideration by defining the functions $|w_{n}^{(cj)}|$ with $j = (1,\cdots,4)$.

\section{Numerical results for two coupled linear chains}
\label{sec:numerics-normal-phase}
\begin{figure*}
\centering
\includegraphics[width=17.8cm]{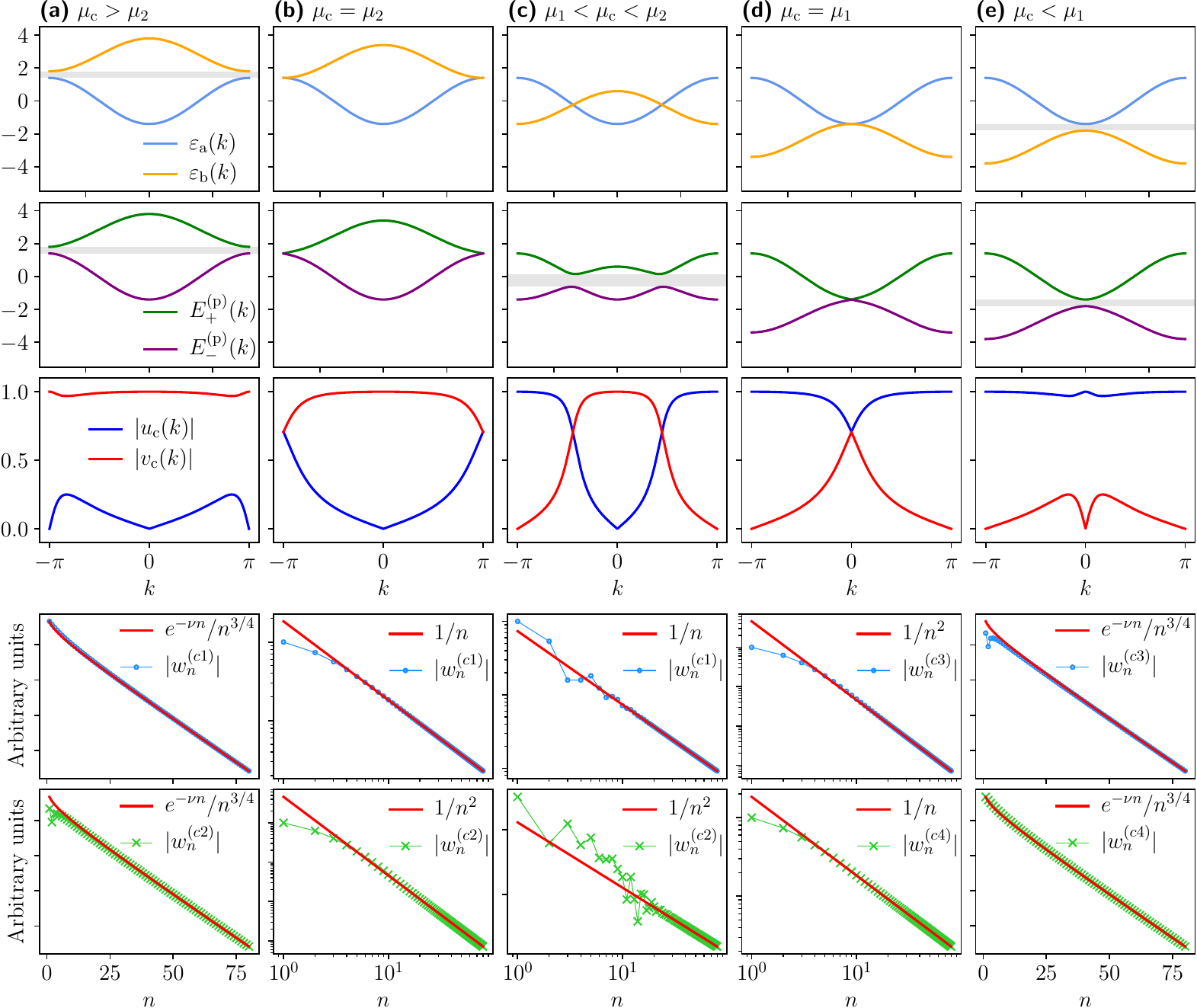}
\caption{Numerical analysis of the two-leg ladder with s-p orbitals, for different regimes of the parameter $\mu_{\rm c}$ of Eq.~\eqref{eq:spchain2x1bis0} across the critical values $\mu_1$ and $\mu_2$ given by Eq.~\eqref{eq:spchain12bis01}. 
             For a given value of  $\mu_{\rm c}$, the panels from \textbf{(a)} to \textbf{(e)} show (from top to bottom): 
             The band dispersions $\varepsilon_{\rm a/b}(k)$ before hybridization [cf. Eq.~\eqref{two-linear-chains-Hamiltonian-201}], the band dispersions $E^{{\rm (p)}}_\pm(k)$ after hybridization [cf. Eq.~\eqref{eq:spchain1}], 
             the factors $|u_{\rm c}(k)|$ and $|v_{\rm c}(k)|$ of Eqs.~\eqref{eq:spchain11}, and the Wannier-like lattice functions $|w^{(c1)}_n|$ and $|w^{(c3)}_n|$ of Eqs.~(\ref{Wannier-like-function-c1}) and (\ref{Wannier-like-function-c3}) 
             which contain the factor $|u_{\rm c}(k)|$ or the Wannier-like lattice functions $|w^{(c2)}_n|$ and $|w^{(c4)}_n|$ of Eqs.~(\ref{Wannier-like-function-c2}) and (\ref{Wannier-like-function-c4}) which contain the factor  $|v_{\rm c}(k)|$ 
             (depending on the parity of the band eigenvectors ~\eqref{eq:spchain12}). 
             In the two top panels the gray shaded stripes highlight the presence of a band gap, while
             in the two bottom panels the red solid lines refer to the expected asymptotic behavior overlapping with the numerical data (blue circles and green crosses). 
             Specifically:
             In panel \textbf{(a)} where $\mu_{\rm c}>\mu_2$, the first trivial phase is identified by non-crossing bands with gaps at $k=\pm\pi$, smooth functions ${\rm sign}(k)|u_{\rm c}(k)|$ and $|v_{\rm c}(k)|$ across the BZ, 
             and an exponential falloff of the Wannier-like lattice functions $|w^{(c1)}_n|$ and $|w^{(c2)}_n|$ like $e^{-\nu n}/n^{3/4}$ (shown in log-linear scale);
             In panel \textbf{(b)} where $\mu_{\rm c}=\mu_2$, the first critical point is identified by the bands touching at $k=\pm\pi$, discontinuous functions ${\rm sign}(k)|u_{\rm c}(k)|$ and $|v_{\rm c}(k)|$ across the BZ, 
             and a power-law falloff of $|w^{(c1)}_n|$ and $|w^{(c2)}_n|$ like $1/n$ and $1/n^2$, respectively (shown in log-log scale);
             In panel \textbf{(c)} where $\mu_1<\mu_{\rm c}<\mu_2$, the non-trivial phase is identified by avoided crossings of the bands with gaps around $k_{\rm x}$ of Eq.~\eqref{eq:positionkcrossingpoint01}, 
             discontinuous functions ${\rm sign}(k)|u_{\rm c}(k)|$ and $|v_{\rm c}(k)|$ across the BZ, and a power-law falloff of the Wannier-like lattice functions $|w^{(c1)}_n|$ and $|w^{(c2)}_n|$ like $1/n$ and $1/n^2$, 
             respectively (shown in log-log scale);
             In panel \textbf{(d)} where $\mu_{\rm c}=\mu_1$, the second critical point is identified by the bands touching at $k=0$, discontinuous functions $|u_{\rm c}(k)|$ and ${\rm sign}(k)|v_{\rm c}(k)|$ across the BZ, 
             and a power-law decay of $|w^{(c3)}_n|$ and $|w^{(c4)}_n|$ like $1/n^2$ and $1/n$, respectively (shown in log-log scale);
             In panel \textbf{(e)} where $\mu_{\rm c}<\mu_1$, the second trivial phase is identified by non-crossing bands with gaps at $k=0$, smooth functions $|u_{\rm c}(k)|$ and ${\rm sign}(k)|v_{\rm c}(k)|$ across the BZ, 
             and an exponential falloff of the Wannier-like lattice functions $|w^{(c3)}_n|$ and $|w^{(c4)}_n|$ like $e^{-\nu n}/n^{3/4}$ (shown in log-linear scale).    
             Data are shown in arbitrary units to ease the comparison among panels with different values of $\mu_{\rm c}$.}
\label{Figure-9}
\end{figure*}

We now numerically analyze in detail across the two QCPs that separate the non-trivial from the trivial topological 
phases for the s-p chain the band dispersions before [cf. Eq.~\eqref{two-linear-chains-Hamiltonian-201}] and after [cf. Eq.~\eqref{eq:spchain1}] hybridization, the factors (\ref{eq:spchain11}) entering the band eigenvectors  \eqref{eq:spchain12}, and the Wannier-like lattice functions (\ref{Wannier-like-function-c1})-(\ref{Wannier-like-function-c4}). Specifically, we show by numerical analysis that, decreasing $\mu_{\rm c}$ across the values $\mu_2=2t_{\rm c}$ and $\mu_1=-2t_{\rm c}$, modifies the band topology of the two connected chains with s- and p-like orbitals, 
driving the system from a trivial to the non-trivial (topological) phase at $\mu_{\rm c}=\mu_2$ and then again to a trivial phase at $\mu_{\rm c}=\mu_1$. 
The numerical results are reported in the various panels of Fig.~\ref{Figure-9}, that we now discuss in detail separately.

\subsection{First trivial region $\mu_{\rm c}>\mu_2$}
\label{sec:region-1}
When $\mu_{\rm c} > \mu_2$ [cf. panels \textbf{(a)} of Fig.~\ref{Figure-9}], the bands $\varepsilon_{\rm a}(k)$ and $\varepsilon_{\rm b}(k)$ do not cross each other, since $\varepsilon_{\rm b}(k)>\varepsilon_{\rm a}(k)$ for all $k\in(-\pi,\pi)$,
with a gap $|\varepsilon_{\rm a}(k)-\varepsilon_{\rm b}(k)|$ that opens at $k=\pm\pi$. 
Accordingly, $\xi_{\rm c}(k)$ of Eq.~(\ref{eq:spchain2x1}) is always negative, such that the factors $|u_{\rm c}(k)|$ and $|v_{\rm c}(k)|$ of Eqs.~(\ref{eq:spchain11}) do not cross each other inside the BZ. 
In this case, the phenomenology is similar to that discussed with reference to panels \textbf{(a)}-\textbf{(c)} of Fig.~\ref{Figure-2} and panels \textbf{(a)} and \textbf{(d)} of Fig.~\ref{Figure-4}. 
Regarding the band eigenvectors, we adopt the same provision used when discussing the one-dimensional p-wave fermionic superfluid, whereby the upper eigenvector components are multiplied by ${\rm sign}(k)$ like in Eqs.~\eqref{eq:spchain12}. 
The Wannier lattice functions of Eqs.~\eqref{eq:numericaltestwannierfunction21} then reduce to $w^{(c1)}_n$ and $w^{(c2)}_n$ of Eqs.~(\ref{Wannier-like-function-c1}) and (\ref{Wannier-like-function-c2}). 

The functions ${\rm sign}(k)\,|u_{\rm c}(k)|$ and $|v_{\rm c}(k)|$ are smooth over the whole the BZ, such that the Wannier-like lattice functions $|w^{(c1)}_n|$ and $|w^{(c2)}_n|$ of Eqs.~(\ref{Wannier-like-function-c1}) and 
(\ref{Wannier-like-function-c2}) falloff exponentially like $e^{-\nu n}/n^{3/4}$ similarly to what occurs in panels \textbf{(a)} and \textbf{(d)} of Fig.~\ref{Figure-4}
(where the exponent $\nu$ depends on the set of parameters entering $\xi_{\rm c}(k)$ of Eq.~(\ref{eq:spchain2x1}) and $\epsilon_{\rm c}(k)$ of Eq.~(\ref{eq:spchain3bis01})). 
This regime of the parameter $\mu_{\rm c}$ thus identifies a \emph{first trivial} (non-topological) region.

\subsection{First quantum critical point at $\mu_{\rm c} = \mu_2$}
\label{sec:region-2}
When $\mu_{\rm c} = \mu_2$ [cf. panels \textbf{(b)} of Fig.~\ref{Figure-9}], the first QCP is reached, which separates the first trivial region from the non-trivial (topological) region. 
The bands $\varepsilon_{\rm a}(k)$ and $\varepsilon_{\rm b}(k)$ of Eq.~\eqref{two-linear-chains-Hamiltonian-201} touch each other at $k=\pm|k_{\rm x}|=\pm\pi$, where $\xi_{\rm c}(k=\pm\pi) = 0$ and also $\tau(k=\pm\pi)=0$. 
Accordingly, the eigenvalues $E^{\rm (p)}_\pm(k)$ of Eq.~\eqref{eq:spchain1} also touch each other at $k=\pm\pi$ such that the band gap closes, while the factors $|u_{\rm c}(k)|$ and $|v_{\rm c}(k)|$ of Eqs.~(\ref{eq:spchain11}) 
\emph{exchange\/} their values at $k=\pm\pi$ where $|u_{\rm c}(k=\pm\pi)| = |v_{\rm c}(k=\pm\pi)|=1/\sqrt{2}$ are nonzero (while keeping their different values at $k=0$).
This ``exchange'' is at the essence of the topological properties of the electronic bands of the material.

As a consequence of this exchange, ${\rm sign}(k)\,|u_{\rm c}(k)|$ and $d|v_{\rm c}(k)|/dk$ are now discontinuous functions at $k=\pm\pi$. These discontinuities cannot be removed without introducing jump discontinuities in these functions. Furthermore, these discontinuities determine the large-$n$ behavior of the Wannier-like lattice functions~(\ref{Wannier-like-function-c1}) and (\ref{Wannier-like-function-c2}), such that the Wannier-like lattice functions $|w^{(c1)}_n|$ and $|w^{(c2)}_n|$ have power-law falloffs like $1/n$ and $1/n^2$, respectively. The numerical results shown in Fig.~\ref{Figure-9} confirm this analytic behavior, since both Wannier-like lattice functions are seen to overlap with the expected trend for sufficiently large $n$.
At $\mu_{\rm c}=\mu_2$ the system thus enters the non-trivial (topological) phase, in a similar way to what occurs at $\mu=2t$ for the p-wave superfluid.

\subsection{Non-trivial region $\mu_1 < \mu_{\rm c} < \mu_2$}
\label{sec:region-3}
When $\mu_1<\mu_{\rm c}<\mu_2$ [cf. panels \textbf{(c)} of Fig.~\ref{Figure-9}], the bands $\varepsilon_{\rm a}(k)$ and $\varepsilon_{\rm b}(k)$ cross each other at $|k_{\rm x}|<\pi$, that is, at $|k_{\rm x}|>\pi/2$ when $\mu_{c}>0$ and
at $|k_{\rm x}|<\pi/2$ when $\mu_{c}<0$. 
Since $\xi_{\rm c}(k_{\rm x})=0$  from Eq.~\eqref{eq:spchain2x1}, the value of $k_{\rm x}$ is also the crossing point of the functions $|u_{\rm c}(k)|$ and $|v_{\rm c}(k)|$ of Eqs.~\eqref{eq:spchain11} 
where $|u_{\rm c}(k_{\rm x})|=|v_{\rm c}(k_{\rm x})|=1/\sqrt{2}$.  
Accordingly, ${\rm sign}(k)\,|u_{\rm c}(k)|$ is odd across the BZ and has a sign discontinuity at $k=\pm\pi$, while $|v_{\rm c}(k)|$ is even with a sign discontinuity in the first derivative at $k=\pm\pi$. The large-$n$ behavior of the Wannier-like lattice functions~(\ref{Wannier-like-function-c1}) and (\ref{Wannier-like-function-c2}) is a power-law falloff like $1/n$ for $|w^{(c1)}_n|$ 
and a power-law falloff like $1/n^2$ for $|w^{(c2)}_n|$. This regime of the parameter $\mu_{\rm c}$ then identifies the \emph{non-trivial} (topological) phase.

\subsection{Second quantum critical point at $\mu_{\rm c} = \mu_1$}
\label{sec:region-4}
When $\mu_{\rm c}=\mu_1$ [cf. panels \textbf{(d)} of Fig.~\ref{Figure-9}], the second quantum critical point between the nontrivial and trivial regions is reached. 
The bands $\varepsilon_{\rm a}(k)$ and $\varepsilon_{\rm b}(k)$ of Eq.~\eqref{two-linear-chains-Hamiltonian-201} touch each other at $k=0$, where $\xi_{\rm c}(k=0) = 0$ and also $\tau(k=0)=0$. 
Accordingly, the eigenvalues $E^{\rm (p)}_\pm(k)$ of Eq.~\eqref{eq:spchain1} also touch each other at $k=0$ such that the band gap closes, while the factors $|u_{\rm c}(k)|$ and $|v_{\rm c}(k)|$ of Eqs.~(\ref{eq:spchain11}) 
exchange their values at $k=0$ where $|u_{\rm c}(k=0)| = |v_{\rm c}(k=0)|=1/\sqrt{2}$ are nonzero (while keeping their values different at $k=\pm\pi$).
Owing to this second exchange, at $\mu_{\rm c}=\mu_1$ the system leaves the non-trivial phase and enters a \emph{second trivial} phase.

In this case, both $|u_{\rm c}(k)|$ and ${\rm sign}(k)\,|v_{\rm c}(k)|$ are discontinuous, in the sense that $|u_{\rm c}(k)|$ has a jump discontinuity in its first derivative at $k=0$, while ${\rm sign}(k)\,|v_{\rm c}(k)|$ has a sign discontinuity at $k=0$. 
These discontinuities affect the Wannier-like lattice functions $|w^{(c3)}_n|$ and $|w^{(c4)}_n|$ of Eqs.~(\ref{Wannier-like-function-c3}) and (\ref{Wannier-like-function-c4}), which are thus expected to falloff for large $n$ like $1/n^2$ and $1/n$, respectively, 
as confirmed by the numerical results shown in Fig.~\ref{Figure-9}. 

\subsection{Second trivial region $\mu_{\rm c} < \mu_1$}
\label{sec:region-5}
When $\mu_{\rm c}<\mu_1$ [cf. panels \textbf{(e)} of Fig.~\ref{Figure-9}], the bands $\varepsilon_{\rm a}(k)$ and $\varepsilon_{\rm b}(k)$ do not cross each other, 
since $\varepsilon_{\rm b}(k)<\varepsilon_{\rm a}(k)$ for all $k\in(-\pi,\pi)$, with a gap $|\varepsilon_{\rm a}(k)-\varepsilon_{\rm b}(k)|$ that opens at $k=0$.
In this case, $\xi_{\rm c}(k)$ is always positive, such that the factors $|u_{\rm c}(k)|$ and $|v_{\rm c}(k)|$ of Eqs.~(\ref{eq:spchain11}) do not cross inside the BZ.
In addition, since $\tau(k)=0$ at $k=(0,\pm\pi)$, it follows that $\epsilon_{\rm c}(k)=\xi_{\rm c}(k)$, $|u_{\rm c}(k)|=1$, and $|v_{\rm c}(k)|=0$ for $k=(0,\pm\pi)$, as shown in Fig.~\ref{Figure-9}.
As a consequence, $|u_{\rm c}(k)|$ is a smooth function of $k$ across the BZ, while $|v_{\rm c}(k)|$ has a discontinuity in the first derivative that can be eliminated by multiplying it by ${\rm sign}(k)$. 
The band eigenvectors are thus suitably identified by multiplying the expressions~\eqref{eq:spchain12} by ${\rm sign}(k)$, such that $|u_{\rm c}(k)|$ is an even and ${\rm sign}(k)\,|v_{\rm c}(k)|$ is an odd function of $k$, and
the relevant Wannier-like lattice functions are again $w^{(c3)}_n$ and $w^{(c4)}_n$ of Eqs.~(\ref{Wannier-like-function-c3}) and (\ref{Wannier-like-function-c4}).

Going through the same type of arguments as for the one-dimensional p-wave fermionic superfluid, the above smooth behavior of the band eigenvectors reflects itself in an exponential falloff $e^{-\nu n}/n^{3/4}$ of the Wannier-like lattice functions for large $n$, 
because the dominant discontinuities are those arising from the zeros in the complex plane of $\epsilon_{\rm c}(z)$ in the expressions~\eqref{eq:spchain11}, which determine the exponent $\nu$. 
The corresponding numerical results for these lattice functions are shown at the bottom of panels \textbf{(e)} of Fig.~\ref{Figure-9} (blue dots and green crosses, respectively), together with the expected asymptotic behavior (red lines).
Note again the similarity with panels \textbf{(a)} and \textbf{(d)} of Fig.~\ref{Figure-4} for the p-wave superfluid.

\subsection{Shifting of the Wannier functions to interstitial positions}
\label{sec:shift-Wannier-functions}
When calculating the Wannier functions of Eq.~\eqref{corresponding-Wannier-functions}, or the related Wannier-like lattice functions of Eqs.~(\ref{Wannier-like-function-c1})-(\ref{Wannier-like-function-c4}), 
the insertion of the extra factor $e^{- i k/2}$ in the in sum over $k$ in the BZ has the effect of compensating for the jump discontinuities in the eigenvectors (whenever they occur).
As a consequence of this compensation, the spatial falloff of Wannier functions becomes much faster, passing from power-law to exponential, but at the same time the Wannier functions become centered about
\emph{interstitial\/} positions (that is, intermediate between sites $n$ and $n \pm 1$) instead than at the lattice sites $n$.
This important feature signals the general tendency of topological materials for building up a ``surface state'' when the bulk system has non-trivial topological band structure, a result that is referred to as the ``bulk-boundary (or bulk-edge) correspondence'' \cite{Bernevig-2013}.
These features were originally pointed out in Ref.~\cite{Strinati-1978}, where the tails of the Wannier functions and their symmetry properties (including the lattice or interstitial positions about which these symmetries are bound to occur) 
were explicitly related to the overall role of the singular points in realistic band structures. 
Later on, these features (and, in particular, the location of the Wannier centers) were related to the occurrence of the Berry phase (cf., e.g., Chap.~3 of Ref.~\cite{Vanderbilt-2018}).

\begin{figure}[t]
\begin{center}
\includegraphics[width=8.5cm,angle=0]{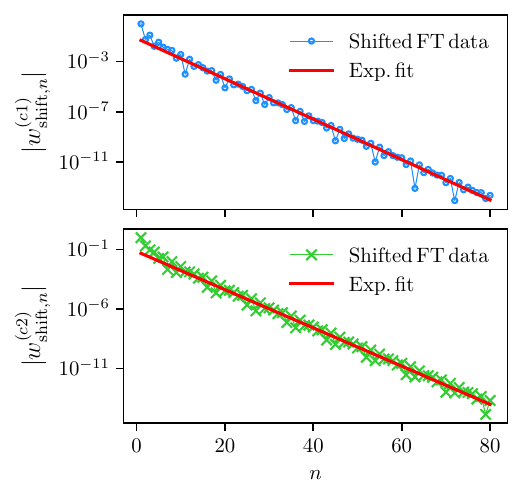}
\caption{The Wannier-like lattice functions, considered in the two bottom rows of panel \textbf{(c)} of Fig.~\ref{Figure-9}, are here recalculated by multiplying by the factor $e^{- i k/2}$ the integrands of
              Eqs.~(\ref{Wannier-like-function-c1}) and (\ref{Wannier-like-function-c2}) (blue circles and green crosses). 
              The ensuing numerical results are compared with the expected asymptotic exponential behavior (red solid lines).
              Note the log-linear scale.}
\label{Figure-10}
\end{center} 
\end{figure} 

These properties can be readily exemplified in terms of the present analysis of the topological properties of the two-leg ladder with s-p orbitals.
In particular, the power-law falloff reported in the two bottom rows of panel \textbf{(c)} of Fig.~\ref{Figure-9} get transmuted into exponential falloffs when the integrands in
Eqs.~(\ref{Wannier-like-function-c1}) and (\ref{Wannier-like-function-c2}) are multiplied by the factor $e^{- i k/2}$, as shown in Fig.~\ref{Figure-10}.
In this case, the non-trivial topology of the band structures in $k$-space is reconciled with the short-range behavior of the Wannier functions in real space, by enforcing an overall shift of the \emph{whole set\/} of Wannier functions 
by half the lattice space.

\section{Conclusions}
\label{sec:conclude}
In this article, we have analyzed in some detail two simple yet instructive examples which are prototypes of one-dimensional topological materials, when the fermionic systems is either in the superfluid or in the normal phases at zero temperature,
and pointed out a close analogy between the topological properties of these two systems.
In this way, we have provided a pedagogical description of the main features that lie, quite generally, at the basis of the topological properties of solid-state materials, with the aim of making them accessible to a wide as possible audience.
In particular, we have identified the key feature of the transition from the trivial to the non-trivial (topological) phases, when a ``folding'' of the two components of the eigenvectors occurs, like when passing from panels \textbf{(c)} to panels \textbf{(e)} of Fig.~\ref{Figure-2}
(or, equivalently, from panels \textbf{(a)} to panels \textbf{(c)} of Fig.~\ref{Figure-9}).
Pictorially, this folding may be considered as the quantum analog in the present context of the topological transition from an ordinary rubber band to a M\"{o}bius strip.

In a sense, in the present article we have proceeded back and retraced the track of the pioneering analysis of Ref.~\cite{Strinati-1978}, about the role of the singular points in the band structure and the related behavior of the tails of the Wannier functions,
which preceded by a few years the introduction of the Berry phase \cite{Berry-1984}.

Yet, it is evident that, after having fully mastered the subtleties of the non-analytic behavior of the Bloch eigenfunctions that give rise to the non-trivial topological properties of the materials (like the “bulk-boundary correspondence”
related to the shift to interstitial positions of the centers of the Wannier functions), advanced mathematical methods are anyhow required for determining the groups of bands that are topologically non-trivial with the associated non-exponential falloff of the Wannier functions \cite{Bernevig-2017}.
To this end, sophisticated mathematical tools have been developed, together with efficient and fully automated algorithms that diagnose the nontrivial band topology, lead to identify new topological phases, and allow for extensive experimental exploration of topological properties.
These tools have eventually led to extensive and rather complete catalogues of topological electronic materials \cite{Zhang-2019,Vergniory-2019,Tang-2019}.

Besides the development of these general tools for a theoretical identification of topological materials, general methods to directly detect and characterize quantum topological states in practice would also be required.
A recent proposal, that utilizes the interference from topological defects, could be a possible candidate in this sense \cite{Akkermans-2025}.
Or else, another recent proposal focuses on how to reveal the topological properties of a one-dimensional p-wave fermionic superfluid in terms of a suitable arrangement of an ultra-cold Fermi gas \cite{Ferlaino-2025}.

\vspace{0.2cm}
\textit{Data availability statement} -- The data that supports the findings of this study are available within the article.

\vspace{0.2cm}
\textit{Conflict of interest} -- The authors have no conflicts to disclose.

\end{document}